\documentclass[a4paper,twoside,10pt]{article}

\usepackage[utf8]{inputenc}  
\usepackage[T1]{fontenc}   
\usepackage{times}
\usepackage{amsmath,amssymb,amsthm,euscript,pifont}
\usepackage{mathrsfs}
\usepackage{fullpage}
\usepackage{graphicx}
\usepackage{subfig}
\usepackage{caption}
\usepackage{bbm}
\usepackage{tikz}
\usepackage{authblk}

\usepackage{color}
\definecolor{Chocolat}{RGB}{93,57,57}

\usepackage[pagebackref,colorlinks=true]{hyperref}
\hypersetup{
    linkcolor = {red}, 
    citecolor = {blue}
    }

\usepackage{xspace}
\usepackage{algorithmic, algorithm}

\newcommand{\er}{\mathbb R}
\newcommand{\nat}{\mathbb N}

\newcommand{\av}[1]{{{\displaystyle{\langle} {#1}\displaystyle{\rangle}}}}
\newcommand{\Ee}{\displaystyle {\mathbb E}}
\newcommand{\Var}{{\mathbb V}\!\textrm{ar}}

\newcommand{\FF}{{\mathcal F}}
\newcommand{\Pp}{{\mathbb P}}

\newcommand{\Xx}{{\rm X}}
\newcommand{\x}{{\rm x}}
\newcommand{\nd}{n_{\mathcal{D}}}

\newcommand{\Uu}{{\rm U}}
\newcommand{\pp}{{\mathscr P}}

\newcommand{\NN}{{\mathscr N}}

\newcommand{\Dd}{\mathcal D}
\newcommand{\tke}{{k}}
\newcommand{\ustar}{u_{*}}

\newcommand{\CC}{\mathcal C}
\newcommand{\ds}{\displaystyle}

\newcommand{\GG}{{U_{\textrm{ext}}}}
\newcommand{\numesp}[1]{{{\displaystyle{\langle}\!\>\!\!{\langle} {#1}\displaystyle{\rangle}\!\>\!\!{\rangle}}}}

\newcommand{\W}{\mathcal{W}}

\usepackage{dsfont}
\def\ind{{\mathds{1}}}

\newcommand{\nacelle}{\textit{nacelle}}
\newcommand{\blades}{\textit{blades}}
\newcommand{\Cblades}{\mathcal{C}_{\textit{blades}}}
\newcommand{\Cnacelle}{\mathcal{C}_{\textit{nacelle}}}
\newcommand{\iin}{\textrm{in}}
\newcommand{\out}{\textit{out}}
\newcommand{\Fmast}{f_{\textit{mast}}}
\newcommand{\Fnacelle}{f_{\textit{nacelle}}}
\newcommand{\dxmill}{\Delta x}
\newcommand{\Urel}{{\mathbf{U}}_{\textrm{relat}}}
\newcommand{\urel}{{U}_{\textrm{relat}}}
\newcommand{\Uinf}{{\mathbf{U}}_{\infty}}
\newcommand{\uinf}{U_{\infty}}
\newcommand{\lm}{\ell_{\textrm{m}}}
\newcommand{\zm}{z_{\textrm{mirror}}}
\newcommand{\zl}{z_{\ell_{\textrm{m}}}}
\newcommand{\zc}{z_{\textrm{c}}}
\newcommand{\xc}{x_{\textrm{c}}}
\newcommand{\yc}{y_{\textrm{c}}}

\newtheorem{remark}{Remark}[section]

\numberwithin{equation}{section}

\title{Modeling the wind circulation around mills with a {L}agrangian stochastic approach}

\author[1]{Mireille Bossy\thanks{mireille.bossy@inria.fr}}
\author[2]{Jos\'e Espina}
\author[2]{Jacques Morice \thanks{jacques.morice@inria.cl}}
\author[2]{Cristi{\'a}n Paris \thanks{cristian.paris@inria.cl}}
\author[3]{Antoine Rousseau\thanks{antoine.rousseau@inria.fr} }

\affil[1]{Tosca Laboratory, Inria Sophia Antipolis -- M\'editerran\'ee, France}
\affil[2]{Inria, Chile}
\affil[3]{Lemon Laboratory, Inria Sophia Antipolis -- M\'editerran\'ee, France}

\date{\today}

\begin{document}

\maketitle
\begin{abstract}
\noindent This work aims at introducing model methodology and numerical studies related to a {Lagrangian} stochastic approach  applied to the computation of the wind circulation around mills. 
We adapt the Lagrangian \textit{stochastic downscaling method} that we have introduced in \cite{SERRA08} and \cite{ber-bos-chauv-jabir-rous-09} to the atmospheric boundary layer and we introduce here a Lagrangian version of the \textit{actuator disc} methods to take account of the mills. We present our numerical method and numerical experiments in the case of non rotating and rotating actuator disc models.  First, for validation purpose we compare some numerical experiments against wind tunnel measurements. Second we perform some numerical experiments at the atmospheric scale  and present some  features of our numerical method, in particular  the computation of  the probability distribution of the wind in the wake zone, as a byproduct of the fluid particle model and the associated PDF method. \end{abstract}
 

\section{Introduction}\label{sec:Intro}

Modeling the flow through wind turbines and wind farms is  a research area of growing importance with the fast and worldwide development of installed wind farms. 
Therefore there exists a wide variety of approaches that  combine atmospheric computational fluid dynamics methods (CFD) with wake models  (from actuator disc models  to  full rotor computations, see eg. S{\o}rensen and Myken \ \cite{sorensen-myken_1992}, Hallanger and Sand \cite{hallanger-sand_2013}, Bergmann and Iollo  \cite{bergmann-iollo_2012}). 

Most popular atmospheric boundary layer computations are based on Reynolds averaged Navier-Stokes (RANS) turbulence models and large eddy simulation (LES) approaches.  
PDF methods, based on  stochastic Lagrangian models, constitutes an interesting alternative (see the discussions  in Pope \cite{pope_2000}) that have been not yet fully developed in the case of atmospheric boundary layer modeling but are mostly used  for reactive flows because this approach does not necessitate to approximate the reaction terms (Haworth \cite{haworth_2010}, Minier and Peirano \cite{minier-peirano_2001}). Among the reasons that can explain this phenomena, we can first rise the fact that such kinds of models handle nonlinear stochastic differential equations that necessitate a background on stochastic calculus, rather than classical  PDE analysis.  Second, the development of numerical solvers, based on stochastic particles approximation, requires the design of a hybrid Lagrangian/Eulerian algorithm from scratch.  However, it is worth mentioning PDF methods are computationally inexpensive and allow to refine the space scale without any numerical constraints. 

In recent works, some of the authors developed  modeling numerical frameworks for the downscaling problem in meteorology (see \cite{SERRA08},\cite{ber-bos-chauv-jabir-rous-09}).  An algorithm called {\it Stochastic Downscaling Method} (SDM) is currently under validation when coupled with a coarse resolution wind prediction, provided thanks to classical numerical weather prediction (NWP) solvers. Quite a few widely used predictive numerical solvers (such as the weather research \& forecasting model (WRF)) are based on a LES approach.  To this purpose, we have developed a branch of SDM that includes atmospheric boundary layer modeling and ground interaction (log law). 

In this paper, we present  our methodology to introduce wake models (typically actuator disc models) in the SDM solver. We also present  the related numerical experiments  on wake computation, in a classical log law context. 

We emphasize the  fact that the basis of  stochastic Lagrangian models consists in describing the stochastic dynamics of a fluid particle's state variables  $(\Xx_t, \Uu_t)$,  under an appropriated probabilistic space $(\Omega, \FF, \Pp)$ equipped with the expectation operator  $\Ee$. The  Lagrangian approach allows to define the Eulerian average of the  velocity, classically represented by the  bracket operator  $\av{\Uu}$ in the RANS approach or $\widetilde{U}$ in the LES approach, as the mathematical  conditional expectation\footnote{We consider here only the case of constant mass density, for the sake of clarity. In other cases a scalar state variable is introduced such as temperature, that weights the  conditional expectation operator.} of the particle velocity $\Uu_t$ knowing  its position  $\x\in \er^3$
\begin{equation}\label{euler:generic-U}
\av{\Uu} (t,\x):=\Ee\left[\Uu_{t}| \Xx_{t}=\x\right],
\end{equation}
and more generally, for any function $f$,  
\begin{equation}\label{euler:generic-fU}
\av{f(\Uu)} (t,\x):=\Ee\left[f(\Uu_{t})| \Xx_{t}=\x\right]. 
\end{equation}
Equivalently, in term of PDF\footnote{Probability Density Function} modeling approach (see \cite{pope_2000} for further details) the Eulerian density at time $t$ and at a given $x$ is identified with the conditioned Lagrangian density, knowing the event $\{X_t=x\}$. 
  
The connections between RANS/LES models and stochastic Lagrangian models are realized via a specific design of the stochastic equation coefficients for $(\Xx_t, \Uu_t)$ allowing to  reproduce $k-\varepsilon$ and Reynolds-stress models (see the review on Lagrangian modeling  in Pope \cite{pope_94} and the discussion in \cite{ber-bos-chauv-jabir-rous-09}).  In particular such PDF models can be used to reconstitute subgrid information of fluctuations in LES modeling. 

\medskip
After a short description of the SDM modeling, mathematical framework, and numerical method in Section~\ref{sec:sdm}, we introduce our  Lagrangian version of the actuator disc modeling in Section~\ref{sec:actuator}.   For validation purpose, in Section~\ref{sec:validation}  we compare some numerical experiments against wind tunnel measurements.   Section~\ref{sec:simulations} is devoted to some numerical experiments at the  atmospheric scale. To assess the mill impact in the flow, SDM simulations are run in both experiments with  the two turbine models presented in Section~\ref{sec:actuator}, namely: the {\textit{non-rotating actuator disc}} model, and the {\textit{rotating actuator disc}} model. 

\section{Stochastic downscaling methodology}\label{sec:sdm}

This section is devoted to the description of the \textit{Stochastic Downscaling Method}, in the framework of turbulence modeling of an incompressible flow in the neutral case (constant mass density). 

Consider the computational domain $\Dd$ as an open bounded subset of $\er^{3}$.
In order to model the flow in $\Dd$, we consider a couple of stochastic processes $(\Xx_t,\Uu_t)$
that respectively describe the location and the velocity of a generic fluid-particle. The evolution of $(\Xx_t,\Uu_t)$
 is governed by the following generic system of stochastic differential equations (SDEs):

\begin{subequations}\label{eq:generic}
\begin{align} 
d\Xx_{t} = & \Uu_{t} dt, \label{eq:generic-X}\\
d\Uu_{t} =& \left(-\frac{1}{\varrho}\nabla_{\x}\langle
\pp \rangle(t,\Xx_{t})\right)dt
-G(t,\Xx_{t})
\Big(\Uu_{t} - \langle \Uu \rangle(t,\Xx_{t})\Big)dt 
+ C(t,\Xx_{t}) dW_{t}. \label{eq:generic-U}
\end{align}
\end{subequations}
Here $W$ is a standard 3-dimensional Brownian motion, $G$, $C$ and $\GG$ are functions to be defined in accordance with the turbulence underlying model (see Section \ref{sec:ABL} below), and $X(t=0)=\Xx_{0}$, $\Uu(t=0)=\Uu_{0}$ where $(\Xx_{0},\Uu_{0})$ are random variables whose probability law $\mu_{0}$ is given. The parameter $\varrho$ is the mass density of the fluid (which is assumed to be constant).

We supplement the dynamics \eqref{eq:generic},  with generic (Dirichlet) boundary conditions on the mean velocity:
\begin{equation}\label{bc:generic-U}
\av{\Uu}(t,\x):=\Ee\left[\Uu_{t}| \Xx_{t}=\x\right]= \GG(t,\x), 
\end{equation}
when $\x = (x,y,z)$ belongs to  the lateral-boundary  part or top-boundary part of $\partial \Dd$.  The bottom-boundary condition that should account for the surface roughness is described in Section \ref{sec:wall-bc}. 

\subsection{Mathematical framewrok}\label{sec:math}

According to \eqref{euler:generic-U}, the term $\av{\Uu}(t,\Xx_{t})=\Ee\left[\Uu_{t}| \Xx_{t}\right]$ denotes the expected velocity of  the particle conditioned by its position $\Xx_{t}$,  making the equation  \eqref{eq:generic} a nonlinear SDEs in the sense of McKean. This means that a Markovian  solution to \eqref{euler:generic-U} must  be a process $((X_{t},U_{t});\,0\leq t \leq T)$ complemented with the set of its time-marginal laws $\rho(t)$ of  $(X_{t},U_{t})$ at any time $\,0\leq t \leq T$,  which  allows to define (assuming the existence of densities for the marginal laws $\rho(t)$)
\[\Ee\left[f(\Uu_{t})| \Xx_{t}=\x\right]  \]
for any measurable function $f$, as the conditional Lagrangian mean
\[\frac{\int_{\er^3} f(u) \rho(t,x,u) du} {\int_{\er^3}\rho(t,x,u) du}\]
whenever the marginal mas $\int_{\er^3}\rho(t,x,u) du$ is positive. 
In the framework of turbulent modeling this conditional Lagrangian mean is then  identified to the Eulerian mean (see Pope \cite{pope_2000} and the references therein)  

In a series of papers, (see  in particular Bossy and Jabir \cite{bossy-jabir-2015}, Bossy, Jabin, Jabir and Fontbona \cite{bossy-fontbona-al-13}, and references therein), the wellposedness of toy-models version of stochastic Lagrangian models, typically with drift coefficients of the velocity equation expressed as conditional expectation with respect to position,  was initiated. 

The wellposedness of the stochastic process $((X_{t},U_{t});\,0\leq t \leq T)$, for any arbitrary finite time $T>0$, whose time-evolution is given by the McKean-nonlinear  SDE
\begin{equation}\label{eq:NonlinearLangevin}
\left\{
\begin{aligned}
&X_{t} =X_{0} +\int_{0}^{t} U_{s}\,ds,\\
&U_{t} = U_{0}+\int_{0}^{t}B[X_s; \rho(s)] ds + \sigma W_{t}, \mbox{ where }\rho(t) \mbox{ is the  density law of }(X_t,U_t)\mbox{ for all }t \in (0, T],
\end{aligned}
\right. 
\end{equation}
can by found in  \cite{bossy-jabir-2015}, 
where $W$ is a standard $\er^{d}$-Brownian motion, the diffusion $\sigma$ is a positive  constant, and the drift coefficient $(x,\psi) \mapsto B(x,\psi)$  is the mapping from $\Dd\times L^{1}(\Dd\times\er^d)$ to $\er^d$ defined by
\begin{equation}\label{eq:DriftDefinitionEDP}
B[x;\psi]=\dfrac{ \int_{\mathbb{R}^{d}} b(v)\psi(t,x,v)dv}
{\int_{\mathbb{R}^{d}}\psi(t,x,v)dv}
\ind_{\{ \int_{\er^{d}}\psi(t,x,v)dv\neq 0\}}
\end{equation}
where $b:\er^{d}\rightarrow \er^{d}$ is a given bounded measurable function. This definition of the drift $B$ makes the mapping $(t,x) \mapsto B[x;\rho(t)]$ coincides with  $(t,x) \mapsto \Ee [b(U_t) | X_t = x]$ and the velocity equation  rewrites
\[U_{t} = U_{0}+\int_{0}^{t}\Ee[b(U_{s})|X_{s}]ds + \sigma W_{t}\]
or equivalently, using the notation in \eqref{euler:generic-U}
\[dU_{t} = \av{b(U)}(t,X_t) dt + \sigma dW_{t}.\]
Moreover a particle system, based on kernel regression estimator of the conditional expectation $B$  is shown to converge weakly to the model  \eqref{eq:NonlinearLangevin}. 	 The construction of the particle approximation is based on local averaging estimate on a $N$-particle set $(X^i_t,U^i_t, i=1,\ldots,N, t\in[0,T])$ of 
\[\Ee[b(U_t)|X_t=x]  \quad\mbox{ by }\quad  \sum_{i=1}^N \W_{N,i}(x) b(U_t^i).\]
Well-known  propositions for the weights $\W_{N,i}(x)$ are Nadaraya-Watson estimator 
\[\W_{N,i}(x) = \frac{K_\epsilon(x-X^i)} {\sum_{j=1}^N K_\epsilon(x-X^j)},\]
for a well chosen kernel $K_\epsilon(x)=K(\frac{x}{\epsilon})$,  and partitioning estimator
\begin{align*}
\W_{N,i}(x) = \frac{\ind_{\{X^i \in \mathcal{B}_{M,j}\}}}{\sum_{k=1}^N \ind_{\{X^k \in \mathcal{B}_{M,j}\}}},\quad\text{for }x\in
    \mathcal{B}_{M,j}
  \end{align*}
given a $M$-partition $\mathcal{P}_M = \{\mathcal{B}_{M,1}, \mathcal{B}_{M,2},\dots,\mathcal{B}_{M,M}\}$ of the domain. It is worth to notice that the algorithm complexity of a particle system based on kernel estimator is up to $\mathcal{O}(N^2)$ whereas the partitioning estimator version is up to $\mathcal{O}(N)$. We retained this last solution for SDM together with some refinement of Particle-in-cell (PIC) technics (see further details in  \cite{SERRA08, ber-bos-chauv-jabir-rous-09}).

Also a confined version of  \eqref{eq:NonlinearLangevin} by mean of specular reflection  is shown to produce  Dirichlet boundary condition,  as mean no-permeability boundary condition:
\begin{equation*}
\Ee[(U_{t}\cdot \nd(X_t))|X_{t} = x]=\av{U\cdot \nd}(t,x) = 0,~\mbox{for}~dt\otimes d\sigma_{\partial\Dd}
\mbox{-a.e.}~(t,x)\in(0,T)\times\partial\Dd.
\end{equation*}

In \cite{bossy-fontbona-al-13},  for  $\Dd$ equal to the torus $\er/\mathbb{Z}$ (and with $(x)\mbox{mod}\, 1:= x- \lfloor x \rfloor$), a step is made in the wellposedness of  Lagrangian equation with pressure term:  
\begin{subequations}
\label{eq:generic_incompressible_lagrangian}
\begin{align}
&X_{t}=\left\lfloor X_{0} +\int_{0}^{t} U_{s}\,ds\right\rfloor,\quad  U_{t}=U_{0}+\sigma W_{t} -\int_{0}^{t}\nabla_{x}  P (s,X_{s})ds-
\beta\int_{0}^{t}(U_{s} -\alpha \Ee(U_s |X_s))ds  \label{SDE}
\\
& law (X_0,U_0) =\rho_0(x,u)dx\, du, \label{eq:sde_init_cond}
\\
&\Pp(X_{t}\in dx)=dx, \mbox{ for all }t\in[0,T],  \label{eq:mass_constraint}
\end{align}
\end{subequations}
For now on, and under drastic hypotheses on the initial condition law, only analytical solutions of the Fokker Planck equation associated to \eqref{eq:generic_incompressible_lagrangian} is established.  This first step contributes to  analyze the role of the gradient pressure term to guarantee the incompressibility constraint on the Eulerian velocity and constant mass density. 
In the  modeling of turbulent flow, the constraint \eqref{eq:mass_constraint} is indeed formulated  heuristically (see e.g. \cite{pope_94}) by rather imposing some  divergence free property on the flow, which in the case of  system  \eqref{eq:generic_incompressible_lagrangian} would correspond to a divergence free condition on the bulk velocity field:
\begin{equation*}
\nabla_x  \cdot 
 \Ee[U_t |X_t=x] =0.
\end{equation*}
By taking the divergence of a formal equation for the  bulk velocity derived from the Fokker-Planck equation,  and a classical projection argument on the space of divergence free fields, it is then assumed that the field $P$  verifies an elliptic PDE, which in our notation  is written as
\begin{equation}\label{ellipticPDE}
\triangle_x P = -\sum_{i,j=1}^d \partial _{x_i x_j} \Ee\left[U^{(i)}_{t}U^{(j)}_{t}|X_t=x\right]
\end{equation}
(see \cite{pope_2000} for a precise formulation).

\subsection{Generic numerical scheme}\label{sec:scheme}

We present hereafter the numerical discretization of equations~\eqref{eq:generic}. It consists in one main time loop in which we identified three main steps: see Algorithm~\ref{algo:SDM}. The interested reader may refer to \cite{SERRA08, ber-bos-chauv-jabir-rous-09} for additional details. In particular, the link between Lagrangian and Eulerian fields (\textit{i.e.} between particles and mesh) is established thanks to classical \textit{particle-in-cell} (PIC) methods (see Raviart \cite{Raviart85}), which are thus used to compute conditional expectations \eqref{euler:generic-U} and \eqref{euler:generic-fU}.  The domain $\Dd$ is divided in partitioning cells defined  from a Cartesian regular mesh.  We denote $N_p$ the total number of particles in the computation, $N_{pc}$ is the number of particles per cell, that is maintained constant in the time step procedure, by the effect of the mass conservation constraint. 

In the case where the \textit{nearest grid point} method (spline of order 0) is used, any conditional expectation such as \eqref{euler:generic-fU} is computed in each cell $\mathcal{C}(i,j,k)$ thanks to an average value over the $N_{pc}$ particles located in the cell:
\begin{equation}\label{eq:numesp}
\numesp{f(\Uu)}(t,\x_{i,j,k})=\dfrac{1}{N_{pc}}\sum_{p\in\mathcal{C}(i,j,k)}f(\Uu^{p}(t)).
\end{equation}
This approach coincides with the partitioning estimator described at Section  \ref{sec:math}. 

\begin{algorithm}
{\tt \small
\caption{-- Time-Step in SDM}\label{algo:SDM}
\begin{algorithmic}
\WHILE{$t_{0}+ n \Delta t<T_{\text{final}}$}
	\STATE {\bf (1)}  Prediction step: move particles thanks to a partial exponential scheme. 
	\STATE {\bf (2)}  Account for boundary conditions
	\STATE {\bf (3)} Correction step: conservation constraints ensuring constant density and free divergence.
\ENDWHILE
\end{algorithmic}
}
\end{algorithm}

For robustness considerations (see Appendix~\ref{subsec:exposcheme}), we consider an exponential version of the explicit Euler scheme for the prediction Step {\bf (1)}. We propose in Step   {\bf (2)} an original method to confine particles in $\Dd$ according to the following downscaling principle: the inferred Eulerian velocity field satisfies the Dirichlet condition \eqref{bc:generic-U}.\\
At time $t_{n-1}= t_0 + (n-1)\Delta t$, the $N_p$ Lagrangian variables $\left(\Xx_{n-1}^p,\Uu_{n-1}^p\right):=\left(\Xx_{n-1}^{p,N_{p}},\Uu_{n-1}^{p,N_{p}}\right)$ are known, as well as the statistics $\tke_{n-1}$ and $\langle U_{n-1}\rangle$ in each cell $\CC$ of the partition of $$\Dd=\displaystyle{\bigcup_{i=1}^{N_c}}\, \CC_i.$$
At time $t_n$, for each particle $p$:

\begin{description}
\item[Step 1.]  \textit{Prediction:} we compute the following quantities
\begin{itemize}
\item[$\bullet$] The particle position $\widetilde \Xx_n^p = \Xx_{n-1}^p+\Delta t \Uu_{n-1}^p$
\item[$\bullet$] The velocity $\widetilde{\Uu}_n^p$ is calculated applying an exponential Euler scheme to the SDE (see Appendix~\ref{subsec:exposcheme})
\begin{equation}\label{eq:expo}
d\widetilde\Uu_t^p = -G(t_{n-1},  \Xx_{n-1}^p) \left(\widetilde\Uu_{t}^p - \av{\Uu_{n-1}}\right) dt +  C(t_{n-1},  \Xx_{n-1}^p)  dW_t,~t\in [t_{n-1},t_n],
\end{equation}
where $\av{\Uu_{n-1}}$, $ \tke_{n-1}$ and $\varepsilon_{n-1}$ are evaluated in the cell containing $\Xx_{n-1}^{p}$.
\end{itemize}
 If $\widetilde \Xx_n^p \in \Dd$, then set $\Xx_n^p=\widetilde \Xx_n^p$ and ${\Uu}_n^p=\widetilde{\Uu}_n^p$.\\[-0.2cm]
\item[Step 2.]  \textit{Boundary condition:} When $\widetilde \Xx_n^p\notin \Dd$; let $t_{\out}$ be the boundary hitting time after $t_{n-1}$, and $\Xx^p_{\out}=\Xx_{n-1}^{p} 
+ (t_{\out} - t_{n-1}) \Uu_{n-1}^p$ be the hitting position, then the reflected position is set to
\begin{equation}
\Xx_n^p = \Xx^p_{\out} - (t_{n} -t_{\out})\Uu_{n-1}^p. \label{eq:refl_pos}
\end{equation}
In concern of the velocity, we simulate Equation~\eqref{eq:expo} between $t_{n-1}$ and $t_{\out}$ with an exponential scheme to obtain the velocity $\Uu_{t_{\out}^-}^p$. Then, in order to match the boundary conditions, we impose a \textit{jump} on the velocity at $t=t_{\out}$:
\begin{equation} \label{eq:refl_velo}
\Uu_{t_{\out}^+}^p = 2\GG(t_{n-1},\Xx^p_{\out}) - \Uu_{t_{\out}^-}^p.
\end{equation}
We finally compute  $\Uu_n^p$ thanks to the numerical computation of Equation~\eqref{eq:expo} between $t_{\out}$ and $t_n$.

\begin{remark}
In a three-dimensional domain, it may happen that $\Xx_n^p$ written in~\eqref{eq:refl_pos} remains outside the computational domain after the reflection, for instance in the neighborhood of the corners. In this case, the particle is replaced near the outward boundary and the new particle position is
set to
\begin{equation*}
\Xx_n^p = \Xx^p_{\out} + {\bf \gamma},
\end{equation*}
where ${\bf \gamma}$ is a small vector pushing $ \Xx^{p}_{\out}$ back into  $\Dd$. The new velocity $\Uu_n^p$ is unchanged.
\end{remark}

\item[Step 3.] {\it Conservation constraints:}
Once the $N_p$ particles are advanced at time $t_n$,
 \begin{itemize}
\item[$\bullet$] move the particles such that there is exactly the same number $N_{pc}$ of particles per cell to fulfill the mass density constraint. To this aim, we use the so-called triangular transport (see \cite{SERRA08}) which consists in sequentially sorting the particle in each of the three space directions. 
This sequential 1D  rearrangement  corresponds to the solution of an optimal transport problem according to the uniform distribution).  
\item[$\bullet$] compute the new Eulerian quantities $\av{ \widetilde \Uu_n}$, and project the new Eulerian velocity field on the divergence free space. This may be done thanks to the classical
resolution of a Poisson equation for the pressure, with homogeneous
Neumann boundary conditions. 
\end{itemize}
\end{description}

\subsection{A specific Lagrangian model for the atmospheric boundary layer} \label{sec:ABL}
 
\begin{figure}[ht]
	\centerline{\includegraphics[height=5.5cm]{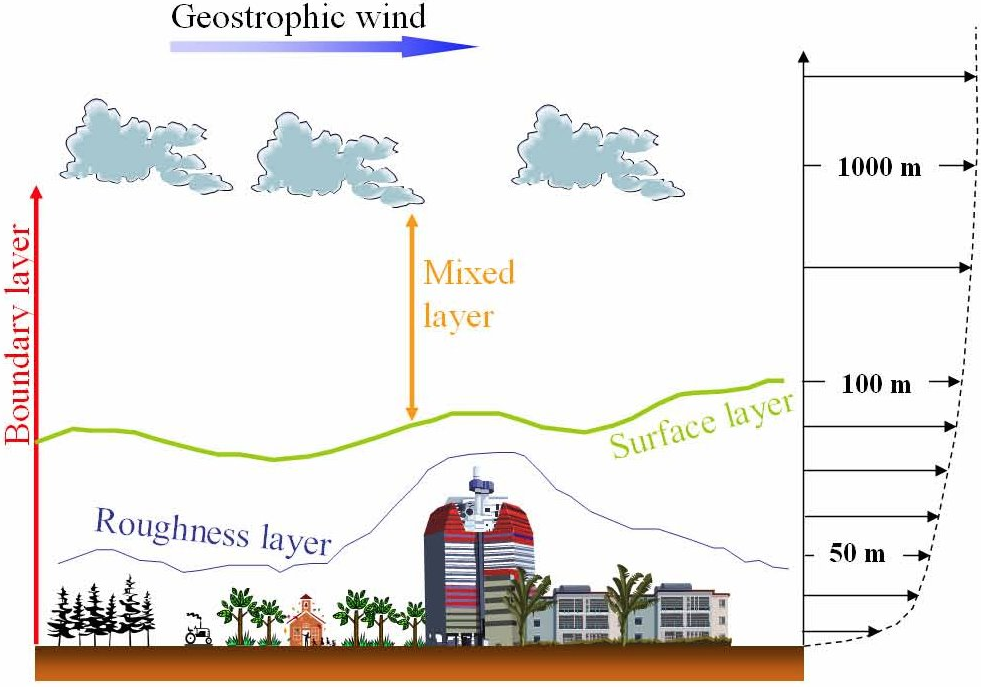}}
	\caption{Schematic view of the atmosperic boundary layer. Courtesy of P. Drobinski \cite{Drobinski2014}.\label{fig:ABL}}
\end{figure}

We consider our computational domain $\Dd$ in the neutral atmospheric boundary layer such as drawn in Figure \ref{fig:ABL}. From floor to top, the height of $\Dd$ is at most  the approximate size of the atmospheric boundary layer, namely $600$ to $1000$ m (in Section \ref{sec:simulations}, we shall perform our numerical simulations for a domain $\Dd$ of height 300 m and 750 m).

We now detail the generic terms $G$, $C$ in  \eqref{eq:generic}--\eqref{bc:generic-U}, and boundary conditions $\GG$ in order to  model the atmospheric boundary layer.
In what follows, all the Eulerian averages $\av{~}$ are in the sense of \eqref{euler:generic-fU}.

\subsubsection*{Turbulence modeling}
We use the classical notation for the velocity components (with numbering or with letters, depending on how it is convenient in the equations) 
$$\Uu_t =(u^{(1)}_t,u^{(2)}_t, u^{(3)}_t) = (u_t,v_t,w_t).$$ 
Also for the components of the instantaneous turbulent velocity, 
$$\Uu_t  - \av{\Uu}(t,\Xx_t) = (u'^{(1)}_t,u'^{(2)}_t, u'^{(3)}_t) = (u_t',v_t',w_t'),$$ 
for the  turbulent kinetic energy (tke),
$$\tke(t,\x)  = \frac{1}{2}\left( \av{u_t'u_t'} + \av{v_t'v_t'} + \av{w_t'w_t'}\right),$$
and  for $\mathcal{P}= \frac{1}{2} ( \mathcal{P}_{11}+\mathcal{P}_{22}+\mathcal{P}_{33})$,  the  rate of  turbulent energy production, with 
\begin{align*}
\mathcal{P}_{ij}:=  - \sum_{k=1}^3 \left(\av{u'^{(i)}u'^{(k)}}
\dfrac{\partial \av{u^{(j)}}}{\partial x_{k}} 
+ \av{u'^{(j)}u'^{(k)}} 
\dfrac{\partial \av{u^{(i)}}}{\partial x_{k}}\right). 
\end{align*}

\medskip
Turbulence models roughly consist in linking the turbulent kinetic energy  $\tke$ and the turbulent energy dissipation  $\varepsilon$. 
In order to account for turbulence effects in the Lagrangian velocity equation \eqref{eq:generic-U}, we define its coefficients as:
\begin{subequations}\label{eq:turbmod}
\begin{align} 
&C(t,\x) = \sqrt{C_{0}\,\varepsilon(t,\x)},\\
&G_{i,j}(t,\x) = -\frac{C_{R}}{2}  \dfrac{\varepsilon(t,\x)}{\tke(t,\x)} \delta_{ij} + C_{2}  \dfrac{ \partial \av{u^{(i)}}}{\partial x_{j}} (t,\x).
\end{align}
\end{subequations}
The tensor $G_{i,j}$ is  related to the isotropization of turbulence production model  (IP model)  that accounts to the Reynolds-stress anisotropies  (see  Durbin and Speziale \cite{durbin-speziale_94} and \cite{pope_2000} ) and $C_0$ is given by 
\begin{eqnarray}\label{IPmodel:C0}
 C_{0} = \frac{2}{3} \left(  C_{R}  + C_2  \frac{\mathcal{P}}{\varepsilon} - 1 \right).
\end{eqnarray}

The turbulent energy dissipation $\varepsilon(t,\x)$ is recovered  via  the turbulent kinetic energy as a parametrization (see   Drobinski et al.  \cite{drobinski-etal_2007}),
\begin{equation}\label{eq:k-eps}
\varepsilon(t,\x)=C_{\varepsilon}\,\dfrac{k^{3/2}(t,\x)}{\lm(\x)}. 
\end{equation}

It is worth to notice that the term $\varepsilon(t,\x)/k(t,\x)$ in the $G_{i,j}$  tensor is $C_{\varepsilon}\,\dfrac{k^{1/2}(t,\x)}{\lm(\x)}$.  Thus when the turbulent kinetic energy $\tke$ vanishes, all the terms in the particles dynamics  (\ref{eq:generic_incompressible_lagrangian}, \ref{eq:turbmod}) stay well defined. As a local model, possibly  forced by dynamical boundary condition as in \cite{SERRA08} or by a log law wind profile  as in the following Sections \ref{sec:validation} and  \ref{sec:simulations}, SDM is mainly pertinent in the turbulent part of the atmosphere.   
However, it can be observed that SDM well reproduces the decay of the turbulent kinetic energy in the boundary layer, and up to the geostrophic height, making vanished the tke $\tke$ at the top, where the flow becomes laminar (see Figure~\ref{fig:atmosBL_main_stat}).

The mixing length $\lm$ can be considered as constant away from the floor (above the surface layer). However, as can be seen in Carlotti \cite{carlotti_2002}, it should be proportional to the vertical coordinate $z$ within the surface layer. A classical choice consists in  a piecewise linear function for $\lm$  proportional to von Karman constant $\kappa$: 
\begin{align}\label{eq:mixinglength}
\lm(z) = \kappa (z - \zl) \ind_{[0,\zl]}(z) + \kappa \zl.
\end{align}

Putting together  \eqref{euler:generic-fU},\eqref{eq:turbmod} with the generic Eulerian average approximation formula \eqref{eq:numesp}, we obtain the following expression for the  turbulent characteristics computed by SDM:
\begin{align}\label{eq:lagrang_tke_sdm}
\begin{aligned}
k(t,\x)&\simeq \frac{1}{2}\left(\numesp{u'u'}(t,\x) + \numesp{v'v'}(t,\x) + \numesp{w'w'}(t,\x)\right).\\
\mathcal{P}_{ij}& \simeq  - \sum_{k=1}^3 \left(\numesp{u'^{(i)}u'^{(k)}}
\dfrac{\partial \numesp{u^{(j)}}}{\partial x_{k}} 
+ \numesp{u'^{(j)}u'^{(k)}} 
\dfrac{\partial \numesp{u^{(i)}}}{\partial x_{k}}\right)\\
G_{i,j}(t,\x) & \simeq -\frac{C_{R}}{2}  C_{\varepsilon}\,\dfrac{k^{1/2}(t,\x)}{\lm(\x)} \delta_{ij} + C_{2}  \dfrac{ \partial  \numesp{u^{(i)}}}{\partial x_{j}} (t,\x).
\end{aligned}
\end{align} 
 
\subsubsection*{Boundary conditions}
As can be seen in Figure \ref{fig:ABL}, our computational domain is bounded from above by the free troposhere where a geostrophic balance can be considered. As a consequence, we shall use Dirichlet boundary conditions at the top of the domain,
\begin{equation}
	\av{\Uu} (t,\x)=U_{G}(t,\x),
\end{equation}
where $U_{G}$ is given, corresponding to the output of a geostrophic model.

The bottom boundary condition should account for the surface roughness and corresponding layer: we incorporate a log law in our model, such as described in the forthcoming section.

Finally, since we want to model one or several mills we propose to use inflow (log law profile) and outflow (free output) boundary conditions for the $(x,y)$ lateral frontiers, as described in Section \ref{sec:simulations} below.

\subsection{Wall-boundary condition}\label{sec:wall-bc}

For the modeling effect of the ground, we borrow and adapt  the {\it particle boundary condition} proposed by Minier and Pozorski \cite{minier-pozorski_99} that aims to reproduce the momentum exchange between the ground and the bulk of the flow.  This method is equivalent to  wall functions approach in classical turbulence models. 

Here, we just resume the main idea in \cite{minier-pozorski_99} which consists in imposing a reflection to the particle trajectories, when it arrive to a given height $\zm$, where $\zm$ is chosen in the logarithmic layer. 

In order to define the reflection of the particles on this mirror face $\{z = \zm\}$, we denote with a  `$\iin$' the inward velocities in  the region $\{z \in [\zm,H]\}$ and with a `$\out$' the outward velocities in  the region $\{z \in [0,\zm]\}$. 

The velocity $U_{\iin}$ is oriented to the top whereas  $U_{\out}$ is oriented to the bottom. A symmetry principle allows to replace any outward particle to its mirror inward particle.  

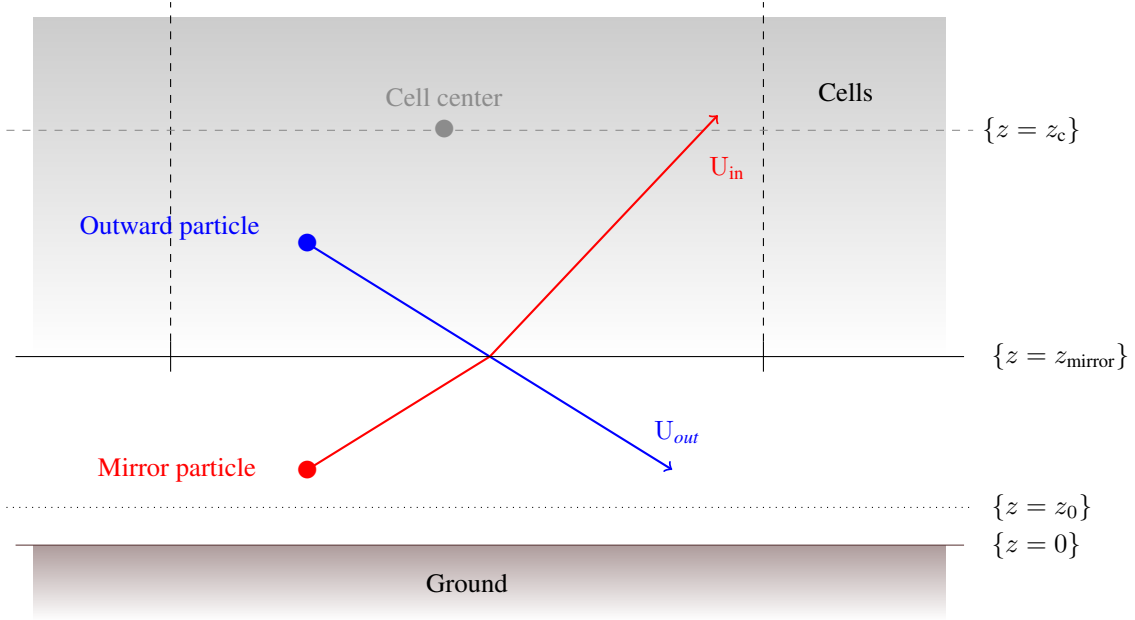
\begin{figure}[ht]
\begin{center}
\begin{tikzpicture}[xscale=1.2,yscale=1]
\newcommand{\zzm}{2.5}
\shade[top color = gray,opacity=.4] (0.0,\zzm) rectangle (10,7);
\draw[black](1.5,2.3) -- (1.5,2.7);
\draw[dashed](1.5,2.7) -- (1.5,7.2);
\draw[gray!90](4.5,5.5) node {{\LARGE$\bullet$}};
\draw[gray!90](4.5,5.7)node[above] {Cell center};
\draw[black](8,2.3) -- (8,2.7);
\draw[dashed](8,2.7) -- (8,7.2);
\draw[black](8.5,6) node[right] {Cells} ;
\draw[gray!90,thin,dashed] (-0.3,5.5) -- (10.3, 5.5);
\draw[black](10.3, 5.5)node[right] {$\{z =\zc\}$};

\draw[blue](0.4,4.2) node[right] {Outward particle} ;
\draw[blue](3,4) node {{\LARGE$\bullet$}} ;
\draw[blue,thick,->] (3,4) -- (7,1);
\draw[blue](6.7,1.5) node[right] {$\Uu_\out$} ;

\draw[red,thick,->] (5,\zzm) -- (7.5,5.7);
\draw[red](7.3,5.0) node[right] {$\Uu_\iin$} ;

\draw[red,thick,-] (3,1) -- (5,\zzm);
\draw[red](3,0.98) node {{\LARGE$\bullet$}} ;
\draw[red](0.6,1) node[right] {Mirror particle} ;

\draw[black,thin,-] (-0.2,\zzm) -- (10.2,\zzm);
\draw[black](10.4, \zzm)node[right] {$\{z =\zm\}$};
\draw[black,dotted] (-0.3,0.5) -- (10.3, 0.5);
\draw[black](10.4, 0.5)node[right] {$\{z =z_0\}$};
\draw[Chocolat,thin,-] (-0.2,0.0) -- (10.2,0.0);

\draw[black](10.4, 0.0)node[right] {$\{z =0\}$};

\shade[top color=Chocolat,opacity=.5] (0.0,0.0) rectangle (10,-1);

\draw[black](4.2, -0.5)node[right] {Ground};
\end{tikzpicture}
\end{center}
\caption{The mirror reflection scheme for the velocity near the ground.\label{fig:reflexion-minier}}
\end{figure}

The vertical component is simply reflected at  $\zm$: 
\begin{align}\label{eq:reflex_in_z}
w_{\iin} = - w_{\out}
\end{align}
whereas the horizontal velocity are lifted in a way that preserves the covariances 
$\av{u'w'}$ and $\av{v'w'}$ in this reflection process:  
\begin{align}
& u_\iin = u_{\out}  -  2 \frac{\av{u'w'}}{\av{w'^{2}}}\, w_{\out}, \label{eq:reflex_in_x}\\
& v_\iin = v_{\out}  -  2 \frac{\av{v'w'}}{\av{w'^{2}}}\, w_{\out}. \label{eq:reflex_in_y}
\end{align}
It remains to impose the covariances at the ground.

In  \cite{carlotti_2002}, Carlotti describes the method used in the Meso-NH model \cite{nmesoh} for the account of the log law. This method is inspired from the one of Schmid and Schumann \cite{schmidt1989}: the boundary condition for the subgrid covariances $\av{u'w'}$ and $\av{v'w'}$ are fixed to 
\begin{eqnarray}
\label{eq:MESO-NH:uw}
\av{u'w'}(t,x,y) &=& -\left(\frac{\av{u}}{\sqrt{\av{u}^2+\av{v}^2}} \ustar^2 \right)(t,x,y)\\
\label{eq:MESO-NH:vw}
\av{v'w'}(t,x,y) &=& -\left(\frac{\av{v}}{\sqrt{\av{u}^2+\av{v}^2}} \ustar^2\right) 
(t,x,y)
\end{eqnarray}
where the friction velocity $\ustar$\footnote{$\ustar= \left(\av{u'w'}^2 +  \av{v'w'}^2\right)^{1/4}$.}  is computed in each cell at the bottom of the domain using the log law 

\begin{eqnarray}\label{eq:MESO-NH:ustar}
\ustar(t,\xc,\yc)  = \kappa 
\frac{ \sqrt{\av{u}^2(t,\xc,\yc,\zc) +\av{v}^2(t,\xc,\yc,\zc) }  }
{\log\ds\left(\frac{\zc}{z_0} \right)}
\end{eqnarray}
where $\kappa$ is the  von Karmann constant, $z_0$ is the roughness length parameter,   and $(\xc,\yc,\zc)$ is the position of a cell's center, for the cells on the floor. 

We adapt this idea in SDM with the following steps. 
\begin{description}
\item[Step 1.]  Given  $\av{u}(t_n)$ and  $\av{v}(t_{n})$, for any of the cells on the floor,\\
- we compute $\ustar(t_n,\xc,\yc)$ with \eqref{eq:MESO-NH:ustar}\\
- we compute $\av{u'w'}(t_n,\xc,\yc)$ and $\av{v'w'}(t_n,\xc,\yc)$ using \eqref{eq:MESO-NH:uw} and \eqref{eq:MESO-NH:vw}. 

\item[Step 2.]  For the particle boundary condition at $\zm$, we localize the particle crossing the interface in the cell of center $(\xc,\yc,\zc)$; we use $\av{u'w'}(t_n,\xc,\yc)$ and  $\av{v'w'}(t_n,\xc,\yc)$ to compute the  reflected  velocity of the mirror particle at the interface, using \eqref{eq:reflex_in_z}, \eqref{eq:reflex_in_x} and \eqref{eq:reflex_in_y}.

\end{description}
In \cite{minier-pozorski_99}, the authors propose to fix $\zm$ to  $\frac{3}{5} \zc$. In our simulations in the following sections, we used $\zm = \frac{1}{2} \zc$.

\section{Actuator disc methods in the Lagrangian setting}\label{sec:actuator}

The presence of wind mills is taken into account thanks to additional force terms in the stochastic differential equations that govern the movement of the particles. To this end, equation \eqref{eq:generic-U}  (which governs the time evolution of the velocity $\Uu_{t} = \left(u_{t},v_{t},w_{t}\right)$ of a particle) is modified as follows:
\begin{equation}\label{eq:generic-U2}
\begin{aligned}
d\Uu_{t} =& \left(-\frac{1}{\rho}\nabla_{x}\av{\pp }(t,\Xx_{t})\right)dt \\ 
&-G(t,\Xx_{t})\Big(\Uu_{t} -\av{\Uu}(t,\Xx_{t})\Big)dt 
+ C(t,\Xx_{t}) dW_{t} \\ 
& + f\left(t,\Xx_{t},\Uu_{t}\right)dt + \Fnacelle\left(t,\Xx_{t},\Uu_{t}\right)dt + \Fmast\left(t,\Xx_{t},\Uu_{t}\right)dt .
\end{aligned}
\end{equation}

The term $ f\left(t,\Xx_{t},\Uu_{t}\right)$ represents the body forces that the blades exert on the flow. The supplementary terms $\Fnacelle\left(t,\Xx_{t},\Uu_{t}\right)$ and $\Fmast\left(t,\Xx_{t},\Uu_{t}\right)$ represent the impact of the mill nacelle and mast. In the present work, only the blade and nacelle forces are considered. 

In this section, we discuss how those force terms should be implemented in the Lagrangian setting considered here, in order to recover consistency with classical Eulerian formulations.

Considering the full geometrical description of the blades (that requires a very fine mesh), the force $f$ is a very complex function which encodes the geometry of the blades; however, in this study we are interested in the overall impact of the mills in the flow, and not in the fine geometrical details of the reciprocal interactions. For this reason, and to avoid costly computations, an \textit{actuator disc} approach is used to provide approximations of $f$, with two different levels of complexity:

\begin{itemize}
\item[(a)] Non-rotating actuator disc with uniform loading.
\item[(b)] Rotating actuator disc.
\end{itemize}

In the Actuator Disc approach, each mill is represented as an immersed surface which concentrates all forces exerted by the mill on the flow. A thorough description of this methodology can be found in Mikkelsen \cite{mikkelsen_2003}, and in the books  by Hansen \cite{hansen_2008} and  Manwell,  McGowan and Rogers \cite{manwell-etal_2002}. With different degrees of complexity, it has been applied to wind turbine simulations in  Port{\'e}-Agel,  Lub and Wu \cite{porteagel-etal_2010},  Master et al. \cite{master-etal_2012},  El Kasmi and  Masson \cite{elkasmi-masson_2008}. It has also been used to simulate arrays of turbines in Wu and Port{\'e}-Agel \cite{wu-porteagel_2012}. 

\begin{figure}[!ht]
\centering
\subfloat[Local coordinates \label{fig:coordinates}]{ 
\includegraphics[scale = 0.25]{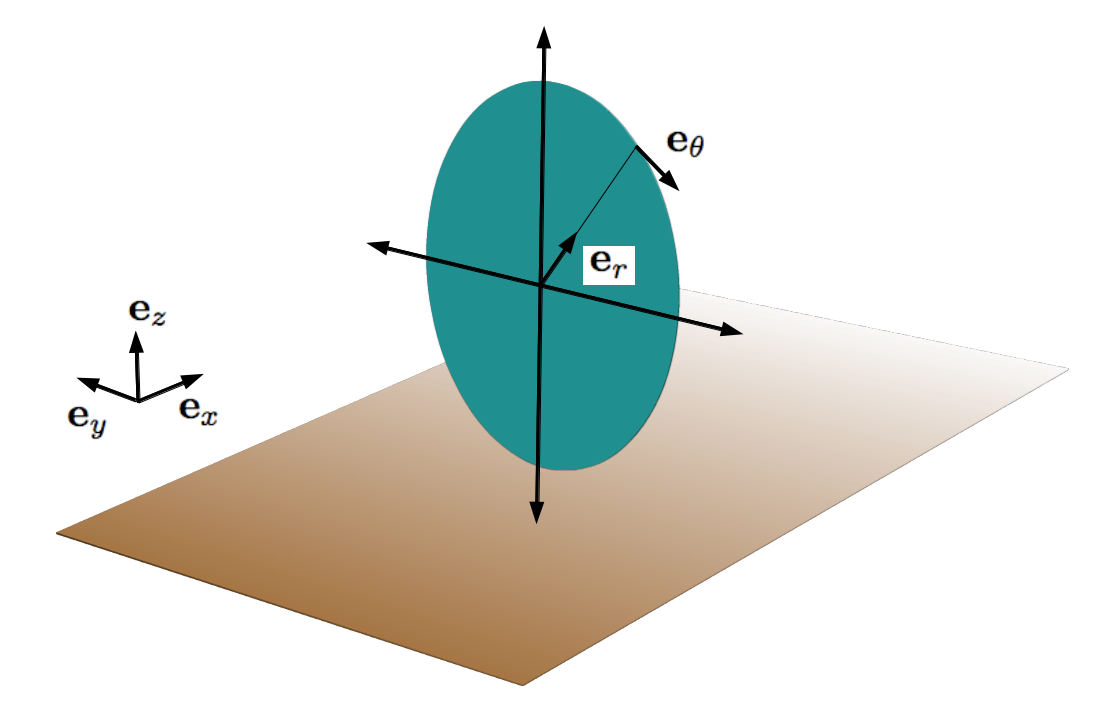}}
\qquad
\subfloat[The cylinder $\mathcal{C}$ \label{fig:cylinder} ]{ 
\includegraphics[scale = 0.25]{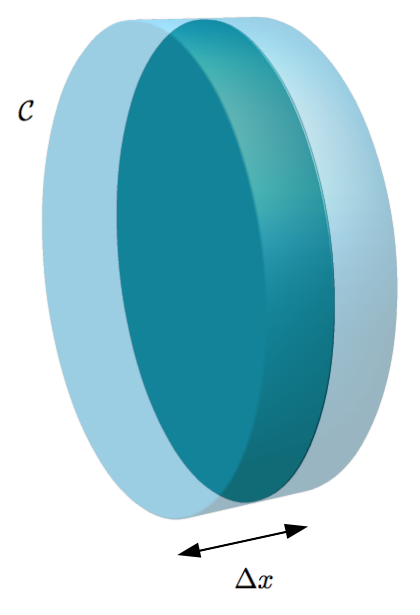}}
\caption{(a) The local reference frame at the actuator disc of the turbine, using cylindrical coordinates. \newline (b) The cylinder $\mathcal{C}$ that extends the actuator disc. Mill forces are applied to particles that lie inside. }
\end{figure}

\begin{figure}[!ht]
	\centering
	\includegraphics[scale = 0.25]{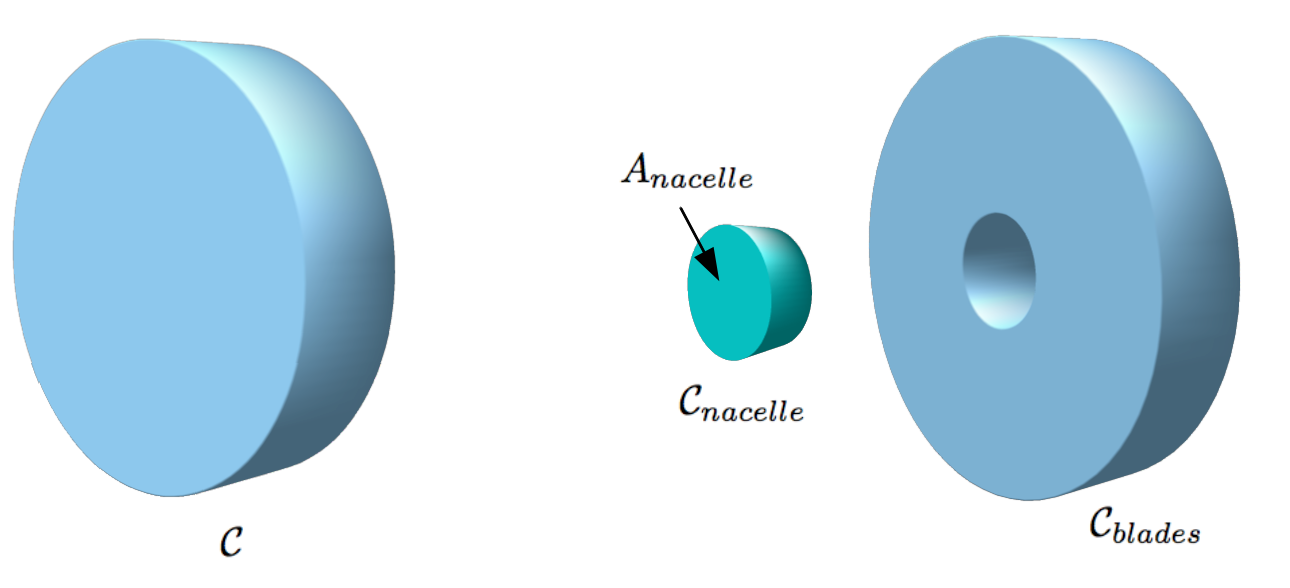}
	\caption{Left: cylinder $\mathcal{C}$ representing the turbine. Right: sub-regions corresponding to the nacelle and blades, viewed by particles. \label{fig:cylinder_nacelle}}
\end{figure}

It is assumed that the flow moves along the positive direction of the $x$ axis, and that the turbine's main axis is aligned with the $x$ axis, so that it faces the wind directly. It is convenient to define local reference frame of cylindrical coordinates centered at the hub of the turbine, with basis vectors ${\bf{e}}_{x}$, ${\bf{e}}_{r}$ and ${\bf{e}}_{\theta}$ as shown in Figure~\ref{fig:coordinates}. In order to apply the forces to the particles, the actuator disc is expanded to a cylinder $\mathcal{C}$ of depth $\dxmill$, and the forces per unit mass are used to correct the velocities of all particles lying inside $\mathcal{C}$. This cylinder is divided in two regions: $\mathcal{C} = \Cblades \cup \Cnacelle$, as depicted in Figure \ref{fig:cylinder_nacelle}, where $\Cblades$ represents the region occupied by the blades, and $\Cnacelle$ represents the region occupied by the nacelle. For the two models considered here (non-rotating and rotating actuator disc), the force term $f$ is computed for and applied to particles lying inside region 
$\Cblades$; and correspondingly, the term $f_\nacelle$ is computed for and applied to particles lying inside region $\Cnacelle$. The following sections describe the way in which this is done.

In the rest of this section we discuss successively on the force $f$  in the non rotating actuator disc model, and in the rotating actuator disc model. We end  by considering the $\Fnacelle$. 

\subsection{Non rotating, uniformly loaded actuator disc model}\label{subsec:ADNR}

In this model, the turbine force corresponding to the blades is distributed uniformly over the region $\Cblades$, and rotational effects are ignored. 
In the simplest formulation of the model, for a turbine facing a uniform laminar steady state flow, and ignoring the influence of the nacelle, the total thrust force exerted by the turbine is given by an expression of the form 

$$F_{x} = -\frac{1}{2} \rho A C_{T} \uinf^2 {\bf{e}}_{x},$$ 

where $\Uinf$ is the unperturbed velocity far upstream from the turbine's location, $\uinf$ is its norm, $A$ is the surface area of the turbine's disc, $\rho$ is the density of air, and $C_{T}$ is a dimensionless, flow dependent parameter called the \emph{thrust coefficient}. 

\begin{figure}[!h]
	\centering
	\includegraphics[scale = 0.30]{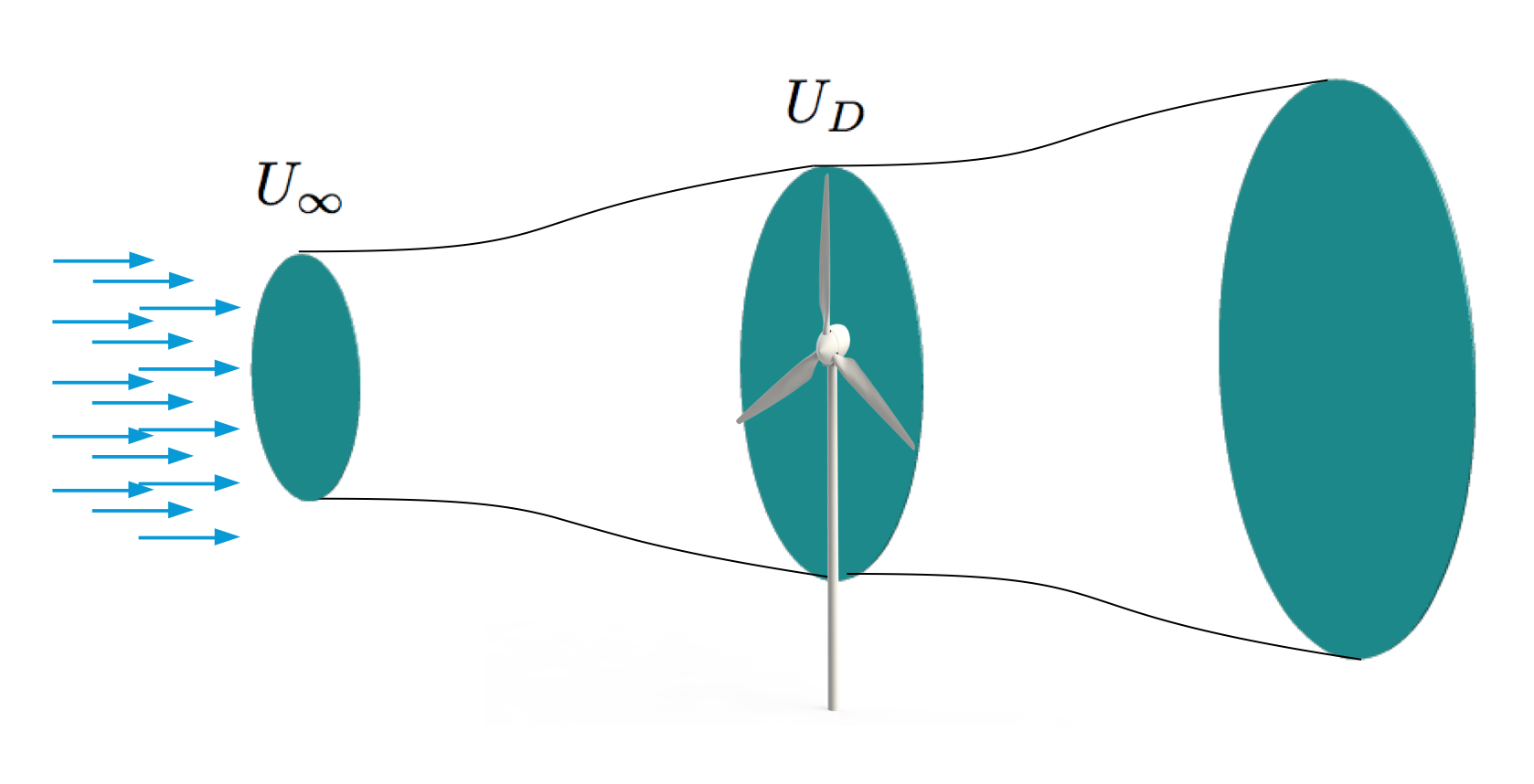}
	\caption{A stream tube passing through the actuator disc of the turbine. Uniform flow is assumed at the far upstream region. \label{fig:streamTube}}
\end{figure}

An elementary deduction of this expression can be found in \cite{hansen_2008} or \cite{manwell-etal_2002}. This deduction, which is essentially one-dimensional, is based on conservation of linear momentum for a stream tube passing through the turbine's disc (see Figure~\ref{fig:streamTube}). The analysis assumes that the turbine faces uniform, inviscid, steady-state flow, and hence there is radial symmetry with respect to the hub of the turbine. Further, a constant loading is assumed at the disc, and thus the velocity field is constant and uniform there. The thrust coefficient $C_{T}$ is specified in terms of the \emph{axial induction factor} $a$,
\begin{equation}\label{eq:inflowFactor} 
a = \frac{\uinf - U_{D}}{\uinf},
\end{equation}
which measures the relative decrease in speed from the far upstream region and the disc region, where the local velocity has magnitude $U_{D} < \uinf$ (see Figure~\ref{fig:streamTube}). 

As in R{\'e}thor{\'e} et al. \cite{rethore-etal_2009}, the local velocity magnitude $U_{D}$ (whose exact formula is yet to be specified) is used instead of $\uinf$ by combining \eqref{eq:inflowFactor} with the thrust force expression that becomes 
 \begin{equation}\label{eq:force2}
F_{x} =  -2\rho \dfrac{a}{(1-a)} A   U_{D}^2   {\bf{e}}_{x}, 
\end{equation}
and 
\begin{equation}\label{eq:thrustCoeff} 
C_{T} = 4a(1-a). 
\end{equation}

In order to adapt this thrust force model  to particles, the disc is extended to a cylinder $\mathcal{C}$ of length $\dxmill$ and mass $\rho A \dxmill$ (see Figure~\ref{fig:cylinder}), which is subdivided in the two regions $\Cblades$ and $\Cnacelle$, as depicted in Figure~\ref{fig:cylinder_nacelle}. In both cases, the force is assumed to be uniformly distributed along the additional dimension. In the present section, we restrict ourselves to the adaptation of the model to the region $\Cblades$. The force per unit mass inside region $\Cblades$ is then given by:

\begin{equation}\label{forceDensity}
f_{x} = -\frac{1}{\dxmill} \frac{2a}{1-a}U_{D}^2 \ind_{\{x \in \Cblades\}}  {\bf{e}}_{x}.
\end{equation}

As soon as we have specified how $U_{D}$ is computed for each individual particle, from  \eqref{forceDensity} follows Algorithm~\ref{algo:nonrotating}, for a given time step of length $\Delta t$ beginning at time $t_{n}$, and a given particle with position $\Xx_{t_{n}}$ and velocity $\Uu_{t_{n}}$ at time $t_{n}$. 
\begin{algorithm}[h!]
{\tt \small
\caption{-- Update the Lagrangian velocity with  thrust force in the non rotating actuator disc model. \label{algo:nonrotating}}
\begin{algorithmic}
\STATE {\bf{PRESTEP}} Compute the mean local speed at the disc $U_{D}^{\textrm{\tiny(Lagrangian)}}$ 
\IF{$\Xx_{t_{n}} \in \Cblades$} \STATE $\Uu_{t_{n+1}} \mapsto \Uu_{t_{n+1}} - \frac{1}{\Delta x} \frac{2a}{1-a}\left(U_{D}^{\textrm{\tiny(Lagrangian)}}\right)^2  \; {\bf{e}}_{x}\;\Delta t.  $ \ENDIF
\end{algorithmic}
}
\end{algorithm}

Since  \eqref{forceDensity} is derived from a laminar one-dimensional analysis, its generalization to turbulent shear flow requires $U_{D}^{\textrm{\tiny(Lagrangian)}}$ to be carefully specified. For a particle with position $\Xx_{t_{n}}$ in $\Cblades$ and velocity $\Uu_{t_{n}}$ at time $t_{n}$, at least three possibilities exist (all equivalent for uniform, laminar, steady-state flows):
\begin{itemize}
\item[(a)] define $U_{D}^{\textrm{\tiny(Lagrangian)}} = \left| u_{t_{n}} \right|$; 
\item[(b)] define $U_{D}^{\textrm{\tiny(Lagrangian)}} = \left| \av{u}  \right| (t_{n},\Xx_{t_{n}})$; 
\item[(c)] compute $U_{D}^{\textrm{\tiny(Lagrangian)}}$ as the magnitude  of average velocity of particles inside cylinder $\Cblades$:
\begin{equation}\label{regularisation1}
U_{D}^{\textrm{\tiny(Lagrangian)}} =  \left| \Ee\left[u_{t_{n}} | \Xx_{t_{n}} \in \Cblades \right] \right|.
\end{equation}
\end{itemize}
If one selects options (a), Equation \eqref{forceDensity} describes the instantaneous force, and not the mean one. 
Given the flow-dependent nature of the axial induction factor $a$, and the fact that it pertains to the whole disc of the turbine, option (c) is adopted, preferably to option (b); the force density then becomes:

\begin{equation}\label{forceDensity2}  
f_{x} = -\frac{1}{\dxmill} \frac{2a}{1-a} \big\lvert \Ee\left[u_{t} | \Xx_{t} \in \Cblades \right] \big\rvert^{2} \ind_{\{\Xx_{t} \in \Cblades\}} {\bf{e}}_{x}.
\end{equation}

\subsection{Rotating actuator disc model}\label{subsec:ADR}

This model is based on a blade element analysis, which gives a description of the blade forces in terms of a set of simple geometrical parameters.  The model assumes that each blade is comprised of tiny pieces (blade elements), each encompassing an infinitesimal length $dr$, which concentrate the relevant forces that the turbine exerts on the flow. It is assumed that these blade elements are independent of one another, in the sense that they do not induce any radial movement on the flow. 

To be precise, consider the reference frame depicted in Figure~\ref{fig:coordinates}, with basis vectors ${\bf{e}}_{x}$, ${\bf{e}}_{r}$, ${\bf{e}}_{\theta}$ along the axial (stream wise), radial and tangential directions, respectively. The corresponding flow velocity components in this frame will be denoted $({\rm U}_{x},{\rm U}_{r},{\rm U}_{\theta})$. We assume that the turbine rotates with angular speed $\omega$, oriented along $-{\bf{e}_{\theta}}$. Consider a blade element located at radius $r$ from the center of the turbine, and a portion of fluid near this blade element. By the model hypotheses, it is assumed that ${\rm U}_{r} = 0$, and thus the flow velocity at this blade element is ${\bf{U}} = {\rm U}_{x} {\bf{e}}_{x} + {\rm U}_{\theta}{\bf{e}_{\theta}}$. Under these conditions, the local relative velocity of the flow with respect to the blade, $\Urel$, is given by: 
\begin{equation}\label{eq:relativeVelocity} 
{\Urel} = {\rm U}_{x}{\bf{e}}_{x} + \left( {\rm U}_{\theta} + \omega\,r \right)  {\bf{e}_{\theta}}. 
\end{equation}

\begin{figure}[!ht]
\centering
\includegraphics[width=0.5\textwidth]{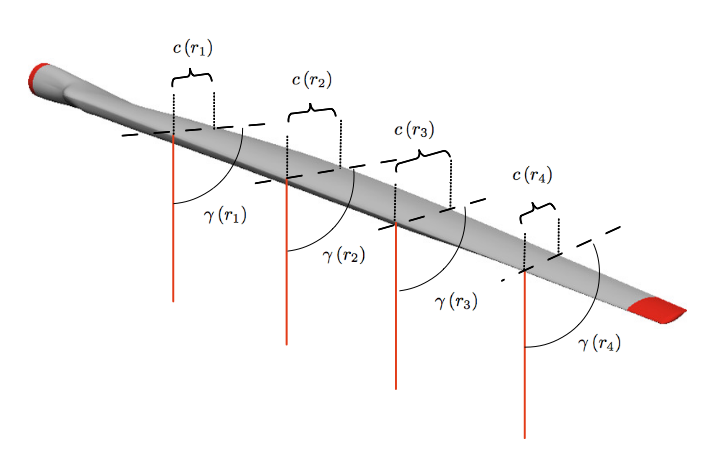}
\caption{A blade, with chord length $c$ and local pitch angle $\gamma$ varying along its radius. Orange (solid) lines lie in the rotational plane of the turbine.} \label{fig:bladeGeometry}
\end{figure}

\begin{figure}[!ht]
\centering
\includegraphics[scale=0.25]{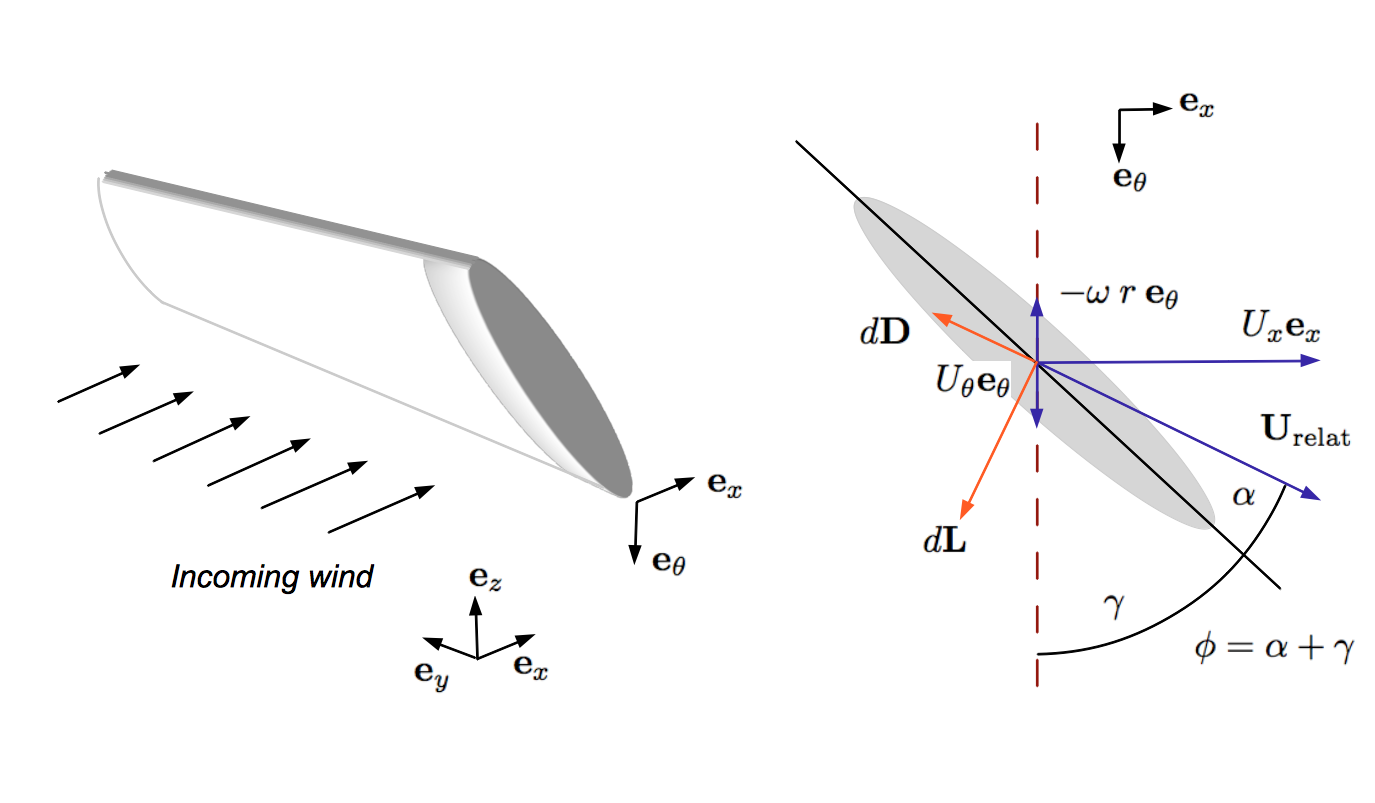}
\caption{Visualisation in the $\left({\bf{e}}_{x} , {\bf{e}}_{\theta}\right)$ plane of a blade element located at radius $r$ from the center of the turbine: relative velocity ${\Urel}$, rotational velocity $-\omega r{\bf{e}}_{\theta}$, local pitch angle $\gamma(r)$ and angle of attack $\alpha$. \label{fig:velScheme} }
\end{figure}

To introduce the blade forces, consider the blade depicted in Figure~\ref{fig:bladeGeometry}. In this model, the actual blade geometry is considered indirectly. The blades themselves are not meshed, but instead are represented by the following information:
\begin{itemize}
\item[{\tiny $\bullet$}] the lift and drag curves corresponding to a given airfoil model of each blade element; 
\item[{\tiny $\bullet$}] the \emph{local chord length} $c(r)$ of the blade at radius $r$, which is the length of the blade element located there (see Figure~\ref{fig:bladeGeometry});
\item[{\tiny $\bullet$}] the \emph{local pitch angle} $\gamma(r)$ of the blade at radius $r$; $\gamma(r)$ is the angle between the chord line of the blade element located at radius $r$, and the rotational plane of the turbine (see Figure~\ref{fig:velScheme}).
\end{itemize}

From this data and ${\Urel}$, two important angles are defined for each blade element (see Figure~\ref{fig:velScheme}):
\begin{itemize}
\item[{\tiny $\bullet$}]  the \emph{flow angle} $\phi$, which is the angle between ${\Urel}$ and the rotational plane of the turbine, and is given by
\begin{equation}\label{eq:flowAngle} 
\phi = \arctan\left(\frac{{\rm U}_{x}}{{\rm U}_{\theta} + \omega\,r}\right),
\end{equation}

\item[{\tiny $\bullet$}] the \emph{angle of attack} $\alpha$, which is the angle between ${\Urel}$ and the main chord line of the blade element, and is given by
\begin{equation}\label{eq:attackAngle} 
\alpha = \phi - \gamma(r).
\end{equation}
\end{itemize} 
 
For the $i$-th blade, \textit{blade element theory} tells us that the total force exerted on the flow by each blade element -- spanning an infinitesimal portion $dr$ of the blade -- is proportional to the square of the norm of $\Urel$ and to the surficial area of the element, which is approximated as $c(r) dr$. The force is divided in two orthogonal components:

\begin{itemize}
\item [1)] A \emph{lift force} ${\bf{dL}}^{(i)}$, which is parallel to $\Urel$
\item [2)] A \emph{drag force} ${\bf{dD}}^{(i)}$, which is orthogonal to $\Urel$
\end{itemize}

The expressions for the magnitudes of these forces, $dL$ and $dD$, read as follows ($\urel$ being the magnitudes of $\Urel$):
\begin{equation}\label{lift1anddrag1} 
\begin{aligned}
dL^{(i)} &= \frac{1}{2} \rho \; \urel^{2}\; c(r) \; C_{L} \; dr,\\
dD^{(i)} &= \frac{1}{2} \rho \; \urel^{2}\; c(r) \; C_{D} \; dr. 
\end{aligned}
\end{equation}

The terms $C_{L}$ and $C_{D}$ are called the \emph{lift coefficient} and \emph{drag coefficient} respectively, and are functions of the angle of attack $\alpha$, as well as the local Reynolds number at the blade. These functions are determined either empirically or via numerical simulations, and are often found in the form of tabulated data. In the present work, two approaches are used to determine $C_{L}$ and $C_{D}$ for a given value of the angle of attack $\alpha$, depending on the mill data we have:
\begin{itemize}
\item[-] numerical interpolation using tabulated data giving the values of both coefficients for a range of possible values of $\alpha$, corresponding to a given airfoil design at a given Reynolds number.  This approach is used in Section \ref{sec:validation}. 

\item[-] direct computation using analytical expressions for the dependence on the angle of attack $\alpha$, corresponding to a given airfoil design at a given Reynolds number. This approach is used in Section \ref{sec:simulations}. 
\end{itemize}

Expressions in \eqref{lift1anddrag1} depend on $i$ as $\urel$, $C_{L}$ and $C_{D}$ are computed locally at each different blade position using one of the methods described above. 

In the local reference frame of the turbine (see Figure~\ref{fig:coordinates}), the components of the differential force along ${\bf{e}}_{x}$ and ${\bf{e_{\theta}}}$ corresponding to the $i$-th blade are, respectively:
\begin{equation}\label{eq:axialForce.and.tangentialForce} 
\begin{aligned}
dF_{x}^{(i)} & = -\left(dL^{(i)} \cos(\phi) + dD^{(i)} \sin(\phi)\right) =  -\frac{1}{2} \rho \; \urel^{2} \; c(r) \; \left(C_{L}\cos(\phi) + C_{D}\sin(\phi) \right) \; dr, \\
dF_{\theta}^{(i)} & = dL^{(i)} \sin(\phi) - dD^{(i)} \cos(\phi) =  \frac{1}{2} \rho \; \urel^{2} \; c(r)\; \left(C_{L}\sin(\phi) - C_{D}\cos(\phi) \right) \; dr.
\end{aligned}
\end{equation}

\begin{figure}[!ht]
	\centering
	\includegraphics[scale = 0.20]{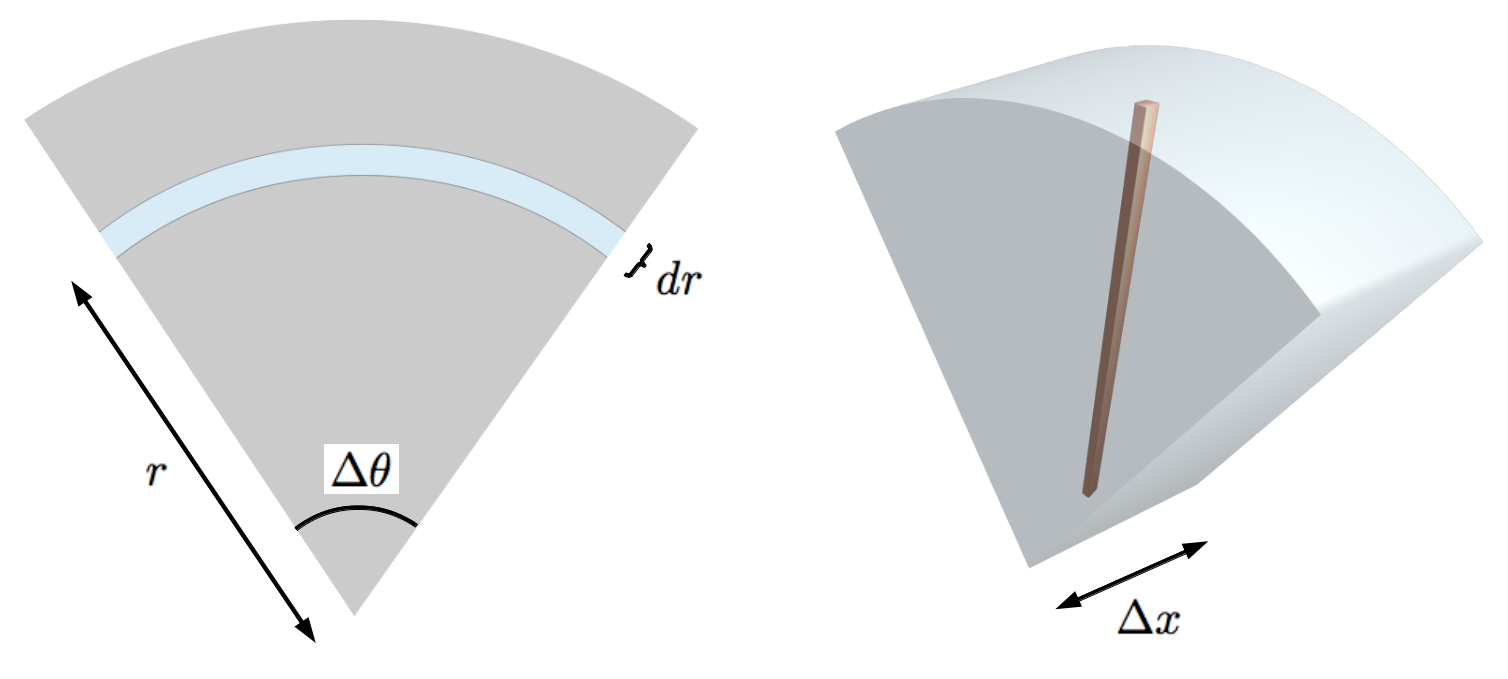}
	\caption{A three dimensional sector corresponding to one blade.}\label{fig:ActuatorLine}
\end{figure}

Now consider a turbine with $N_\blades$ blades. The aim is to use \eqref{eq:axialForce.and.tangentialForce} to obtain expressions for the components of the force per unit mass along ${\bf{e}}_{x}$ and ${\bf{e}}_{\theta}$, corresponding to each blade. To do this, the blade in question is expanded a distance $\Delta x$ along the axial direction, and smeared over an angular distance $\Delta \theta$, resulting in the three dimensional region shown in Figure~\ref{fig:ActuatorLine}. To a blade element located at radius $r$ there corresponds an infinitesimal sector of volume $r \Delta x \Delta \theta dr$ and mass $\rho r \Delta x \Delta \theta dr$.  
Dividing the expressions in \eqref{eq:axialForce.and.tangentialForce} by the element sector  mass, one obtains the forces per unit mass at radius $r$ from the center of the wind turbine: 
\begin{equation}\label{eq:axialForce2.and.tangentialForce2} 
\begin{aligned}
f_{x}^{(i)} & = \frac{1}{\rho r \Delta \theta \Delta x}\frac{dF_{x}^{(i)}}{dr} =  -\frac{1}{2r \Delta \theta \Delta x} \urel^{2} c(r) \left(C_{L}\cos(\phi) + C_{D}\sin(\phi) \right), \\
f_{\theta}^{(i)} & =  \frac{1}{\rho r \Delta \theta \Delta x}\frac{dF_{\theta}^{(i)}}{ dr} = \frac{1}{2r \Delta \theta \Delta x} \urel^{2} c(r) \left(C_{L}\sin(\phi) - C_{D}\cos(\phi) \right).
\end{aligned}
\end{equation}

\begin{figure}[!ht]
	\centering
	\includegraphics[scale = 0.25]{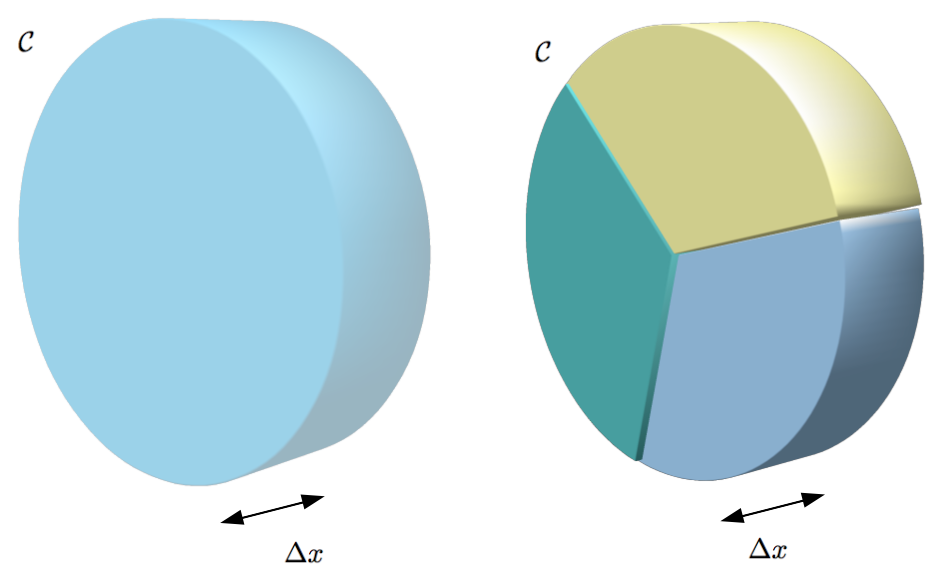}
	\caption{Left: cylinder $\mathcal{C}$ representing the turbine viewed by a particle. Right: decomposition of $\mathcal{C}$ in three  blade sectors. In these figures the nacelle sector $\Cnacelle$ is omitted for simplicity.  \label{fig:cylinder2}}
\end{figure}

Note that expressions in \eqref{eq:axialForce2.and.tangentialForce2} correspond to the force per unit mass of one single blade. In principle, each particle of the simulation will receive at most the force of one blade (the one corresponding to the sector where the particle is), and thus it is necessary to keep track of the positions of each blade (and the corresponding blade sector). 
However, if one is only interested in the global impact of the blades, it is possible to simplify the computations by selecting $\Delta \theta = \frac{2 \pi}{N_\blades}$, where $N_\blades$ is the number of blades of the turbine. In this case, the union of all blade sectors results in the same region $\Cblades$ used in section \ref{subsec:ADNR} (see Figure~\ref{fig:cylinder2}), and the blade positions become irrelevant. 
Expressions  \eqref{eq:axialForce2.and.tangentialForce2} then read as follows:
\begin{equation}\label{eq:axialForce3.and.tangentialForce3} 
\begin{aligned}
f_{x} & = \frac{1}{\rho 2 \pi r  \Delta x} \frac{dF_{x}}{dr} =  -\frac{N_\blades}{4 \pi r \Delta x} \urel^{2} c(r) \left(C_{L}\cos(\phi) + C_{D}\sin(\phi) \right), \\
f_{\theta} &  = \frac{1}{\rho 2 \pi r  \Delta x} \frac{dF_{\theta}}{dr} = \frac{N_\blades}{4 \pi r \Delta x} \urel^{2} c(r) \left(C_{L}\sin(\phi) - C_{D}\cos(\phi) \right).
\end{aligned}
\end{equation}

Note that this is equivalent to summing the contributions of \eqref{eq:axialForce.and.tangentialForce} for all blades, and distributing the resulting force in an annulus of mass $2 \pi \rho r \Delta x dr$ (or equivalently, to choosing $\Delta \theta = 2 \pi$ and summing the contributions of \eqref{eq:axialForce2.and.tangentialForce2} for the $N_\blades$ blades). In this case, since the blades are essentially delocalised within the mill region, their contributions may be considered equal to one another, and the total force at radius $r$ will be given by:
\begin{equation}\label{eq:axialForceTotal.and.tangentialForceTotal}
\begin{aligned}
&dF_{x} = \sum_{i=1}^{N_\blades} dF_{x}^{(i)} = -N_\blades\left(dL \cos(\phi) + dD \sin(\phi)\right), \\
&dF_{\theta} = \sum_{i=1}^{N_\blades} dF_{\theta}^{(i)} =
 N_\blades \left(dL \sin(\phi) - dD \cos(\phi)\right).
\end{aligned}
\end{equation}

Under these assumptions, since the blade characteristics (local pitch, local chord length, lift and drag coefficients) are defined locally, and since the blade positions within $\Cblades$ are indeterminate, we may use the particle positions and Lagrangian velocities  $(\Xx_t,\Uu_t)$  to compute the angle of attack $\alpha$, the local pitch angle $\gamma$ and deduce all the needed information $\Urel$, $c(r)$,  $C_{L}(\alpha)$ and $C_{D}(\alpha)$. From \eqref{eq:relativeVelocity}, we derive the relative velocity $\Urel$ from the instantaneous  particle position and Lagrangian velocity $(\Xx_t,\Uu_t)$ by
\begin{align}\label{eq:urel-intantaneous}
\Urel(\Xx_t,\Uu_t)  = (\Uu_{t}\cdot {\bf{e}_x}){\bf{e}_x} 
+\left( 
(\Uu_{t} \cdot {\bf{e}_\theta})+ \omega\,{r}( \Xx_{t}) 
\right) {\bf{e}_{\theta}}. 
\end{align}
We also compute the flow angle $\phi$ using \eqref{eq:flowAngle}:
\begin{align}\label{eq:phi-intantaneous}
\begin{aligned}
\phi(\Xx_t,\Uu_t) &= \arctan\left(
\frac{(\Uu_{t}\cdot \bf{e}_x)}
{(\Uu_{t}\cdot {\bf{e}_{\theta}}) + \omega\,r( \Xx_{t}) }\right),\\
\alpha(\Xx_t,\Uu_t) &= \phi(\Xx_t,\Uu_t)  - \gamma( r( \Xx_{t})).
\end{aligned}
\end{align}
Then the force added in the Lagrangian velocity Equation \eqref{eq:generic-U2} is  
\begin{equation}\label{eq:axialForce4.and.tangentialForce4} 
\begin{aligned}
f_{x}(t,\Xx_t,\Uu_t) &=-  \ind_{\{\Xx_t \in \Cblades\}} 
\frac{N_\blades}{4 \pi r \Delta x} 
\left(\urel(\Xx_t,\Uu_t) \right)^{2} c(r( \Xx_{t})) 
\left(C_{L}(\alpha)\cos(\phi) + C_{D}(\alpha)\sin(\phi) \right) (\Xx_t,\Uu_t),\\
f_{\theta}(t,\Xx_t,\Uu_t) &=  \ind_{\{\Xx_t \in \Cblades\}} \frac{N_\blades}{4 \pi r \Delta x} 
\left(\urel(\Xx_t,\Uu_t) \right)^{2} c(r( \Xx_{t})) \left(C_{L}(\alpha)\sin(\phi) - C_{D}(\alpha)\cos(\phi) \right)(\Xx_t,\Uu_t).
\end{aligned}
\end{equation}
This results in the numerical Algorithm~\ref{algo:rotating}.

\begin{algorithm}[!h]
{\tt \small
\caption{-- Update the Lagrangian velocity with  thrust force in the rotating actuator disc model.\label{algo:rotating}}
\begin{algorithmic}
\STATE {\bf{PRESTEP}} Compute the relevant geometrical information of the particle at time $t_{n}$, relative to the turbine:
\begin{itemize}
\item the radial position $r\left( \Xx_{t_{n}} \right)$ of the particle (component of $\Xx_{t_{n}}$ along ${\bf{e}}_{r}$)
\item the distance to the rotational plane of the turbine. 
\end{itemize}
With this information, determine if $\Xx_{t_{n}}$ lies inside $\Cblades$ or not.
\IF{$\Xx_{t_{n}} \in \Cblades$} 
	\STATE (1) compute the tangent vector ${\bf{e}}_{\theta}$ corresponding to the projection of $\Xx_{t_{n}}$ on the rotational plane of the turbine; 
	\STATE (2) compute the tangential velocity of the particle, $\Uu_{t_{n}}^{(\tan)}$; that is, its component along ${\bf{e}}_{\theta}$;
	\STATE (3) compute the relative velocity $\Urel$ using \eqref{eq:urel-intantaneous} and the particle velocity components at time $t_{n}$:
	
	$$ \Urel = u_{t_{n}}{\bf{e}}_{x} + \left(  \Uu_{t_{n}}^{(\tan)} + \omega\,r\left( \Xx_{t_{n}} \right) \right) {\bf{e}_{\theta}};$$

	\STATE (4) compute the flow angle $\phi$ using \eqref{eq:phi-intantaneous} and the particle velocity components at time $t_{n}$:
	
	$$ \phi = \arctan\left(\frac{u_{t_{n}}}{\Uu_{t_{n}}^{(\tan)} + \omega\,r\left( \Xx_{t_{n}} \right) }\right); $$
		
	\STATE (5) compute the angle of attack $\alpha$ using \eqref{eq:phi-intantaneous}, with the local pitch computed at the radial position of the particle:
	
	$$ \alpha = \phi - \gamma\left( r\left( \Xx_{t_{n}} \right) \right); $$
	
	\STATE (6) obtain the lift and drag coefficients $C_{L}(\alpha)$ and $C_{D}(\alpha)$ using the airfoil data;
	\STATE (7) apply the axial and tangential forces per unit mass \eqref{eq:axialForce4.and.tangentialForce4} respectively to the particle velocity:
	
$$u_{t_{n+1}}\longmapsto u_{t_{n+1}} -  \frac{N_\blades}{4\pi r  \Delta x} \Urel^{2} c \left(C_{L}\cos(\phi) + C_{D}\sin(\phi) \right) \Delta t,$$
	
$$\Uu_{t_{n+1}}^{(\tan)} \longmapsto \Uu_{t_{n+1}}^{(\tan)} + \frac{N_\blades}{4\pi r  \Delta x} \Urel^{2} c \left(C_{L}\sin(\phi) - C_{D}\cos(\phi) \right) \Delta t. $$
\ENDIF
\end{algorithmic}
}
\end{algorithm}

\subsection{Nacelle model}
In all the simulation results presented in Sections \ref{sec:validation} and \ref{sec:simulations}, we use a simple model for the turbine nacelle. As in  Wu and Port{\'e}-Agel \cite{wu-porteagel_2011}, the nacelle force $\Fnacelle $ is modeled as a permeable actuator disc, but here we adapt the model to the Lagrangian setting. The nacelle is assumed to occupy a cylinder $\Cnacelle$, with frontal area $A_{\nacelle}$ and depth $\Delta x$ (see Figure~\ref{fig:cylinder_nacelle}).  

A particle $(X_t,U_t)$ lying within the nacelle region is applied a force per unit mass of:
\begin{equation}\label{eq:forceDensityNacelle} 
\Fnacelle = -\dfrac{1}{\Delta x}\frac{2 a_{\nacelle}}{1-a_{\nacelle}} \ind_{\{\Xx_t \in \Cnacelle\}} (U_t\cdot {\bf{e}}_{x})^2 {\bf{e}}_{x}.  
\end{equation}

At the same time, the actuator disc models presented in Section \ref{subsec:ADR} are modified to account for the nacelle's presence: the relevant region for the mill forces will not be $\mathcal{C}$, but instead the subregion $\Cblades$ defined as (see Figure~\ref{fig:cylinder_nacelle}):
\begin{equation}\label{eq:redefineC} 
\Cblades =  \mathcal{C} \setminus \Cnacelle.
\end{equation}
In this manner, only particles belonging to $\Cblades$ (and not $\mathcal{C}$) are applied the mill forces. Also, for the Non-Rotating Actuator Disc, the local velocity is estimated considering only particles lying inside $\Cblades$.

\section{Comparison with high-resolution windtunnel measurements}\label{sec:validation}

In this section, the method presented above is tested against wind tunnel measurements performed at the Saint Anthony Falls Wind Tunnel, University of Minnesotta, by Chamorro and Port{\'e}-Agel \cite{chamorro-porteagel_2010}. The data consists of high resolution, hot-wire anemometry wind speed measurements at different downstream positions and heights.  We particularly focus on three key turbulence statistics that are commonly used to characterize wind-turbine  wakes:  the mean and turbulence intensity  profiles of the streamwise velocity,  and  the kinematic shear stress profile  (see e.g. Wu and Port{\'e}-Agel \cite{wu-porteagel_2011}).    

\subsection{Experimental setup}

The Saint Anthony Falls Laboratory wind tunnel consists of two main sections. The air recirculates among these main sections. Turbulence is created using a picket fence, and an adjustable ceiling height allows a zero-pressure gradient boundary layer flow to be created in the main sections. A miniature wind turbine is located in the tunnel, and sensors are placed at different upstream and downstream positions relative to the wind turbine (see \cite{chamorro-porteagel_2010} for details).

For the comparison between Lagrangian simulations and measurements, we used the neutrally-stratified boundary layer experiment. The main characteristics of the neutral boundary layer flow produced in the wind tunnel are summarized in Table~\ref{tab:windtunnel_BL}. 

\begin{figure}[ht]
	\centering
   \includegraphics[width=0.6\textwidth]{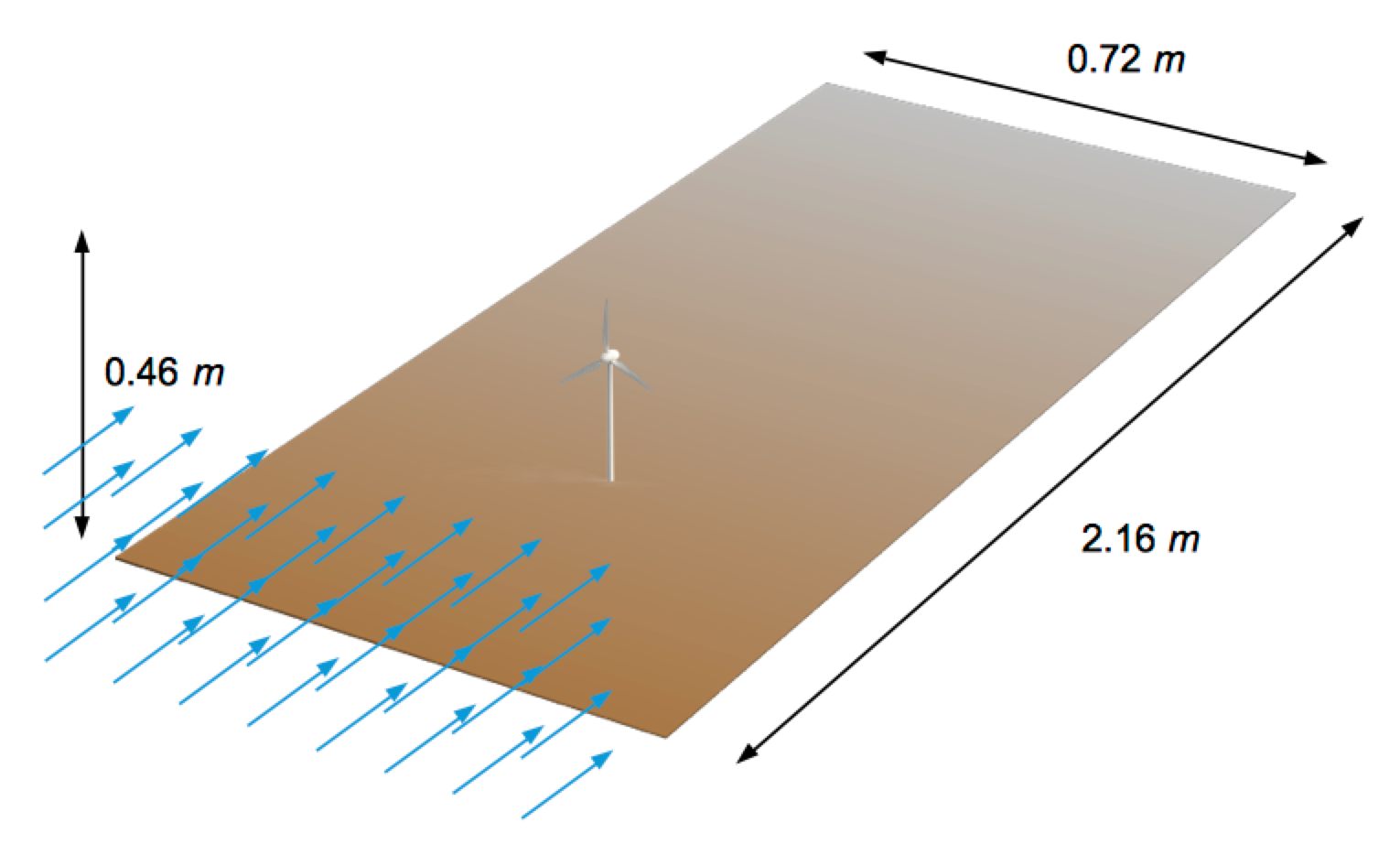}
	\caption{Domain for the wind tunnel  scale simulations. \label{fig:domain2}}
\end{figure}

To test the implementation of mills in the stochastic Lagrangian settings presented before, a one-mill configuration reproducing the Saint Anthony Falls experiment has been used. A single mill has been placed in a rectangular domain spanning $4.32\times 0.72\times 0.46$~meters in the $x$, $y$ and $z$ directions, respectively (Figure~\ref{fig:domain2}). The mill faces atmospheric flow with a log-law profile at the inlet section, that develops moving along the $x$ direction. The main physical and computational parameters of the simulations are detailed in Table~\ref{tab:simudataTunnel}. The miniature three-blade  wind turbine is chosen so as to represent the typical dimensions and tip speed ratios of commercial wind turbines. The main parameters of the wind turbine used in the simulations are listed in Table~\ref{tab:simudataSmallMill}.

\begin{table}[h!]
\centering
\subfloat[Main characteristics of the boundary layer flow \label{tab:windtunnel_BL}]{
\begin{tabular}{| l | l |}
	\hline
	\multicolumn{2}{|c|}{\textbf{Boundary layer characteristics in the wind tunnel}}\\
	\hline
	Boundary layer depth  & 0.46 m \\
	\hline
	Wind speed at the top & 2.8 m\,$\text{s}^{-1}$ \\
	\hline
    Velocity friction $\ustar$ & 0.11  m\,$\text{s}^{-1}$ \\
	\hline
	Roughness length $z_0$ & 0.04 mm \\
	\hline
	\end{tabular}
}\quad
\subfloat[{Main model physical constants\label{tab:windtunnel_PhC}}]{
\begin{tabular}{| l | l |}
	\hline
	\multicolumn{2}{|c|}{\textbf{Model constants}}\\
	\hline
	Rotta constant $C_R$  &  1.8\\
	\hline
	von Karman constant $\kappa$ & 0.4 \\
	\hline
    $C_2$  &  0.60\\
	\hline
	$C_\epsilon$ & 0.068\\
	\hline
	$\zl$ & 0.1 m \\
	\hline
	\end{tabular}
}
\\
\subfloat[configuration of the simulations \label{tab:simudataTunnel}]{
\begin{tabular}{| l | l |}
\hline
\multicolumn{2}{|c|} {\textbf{Simulation parameters}} \\
\hline
Domain size $x$ & 2.16 m  \\
\hline
Domain size $y$ & 0.726 m \\
\hline
Domain size $z$ & 0.42 m \\
\hline
96 cells in $x$ & $\Delta x$ = 0.0225  m\\
\hline
33 cells in $y$ & $\Delta y$ = 0.022 m\\
\hline
84 cells in $z$ &  $\Delta z$ = 0.005 m\\
\hline
Particles per cell & 150 \\
\hline
Final time is 30 s & Time step is  0.03 s \\
\hline
\end{tabular}
}\quad 
\subfloat[parameters of the mill \label{tab:simudataSmallMill}]{
\centering
\begin{tabular}{| l | l |}
\hline
\multicolumn{2}{|c|} {\textbf{Mill configuration}} \\
\hline
Coordinates of the hub: & (0.5 0.36 0.125) m \\ 
\hline
Hub height & 0.125 m  \\
\hline
Radius & 0.075 m \\
\hline
Nacelle radius &  0.01 m \\ 
\hline
Rotational speed & 112.0 $\text{rad.s}^{-1}$ \\
\hline
Inflow factor of the nacelle & $a_\nacelle$ = 0.4 \\
\hline
\end{tabular}
}
\caption{Main parameters of the wind tunnel simulations.} \label{tab:windtunnel}
\end{table}

To assess the impact of the mill in the flow, Lagrangian simulations are run with the  \textit{rotating actuator disc} turbine model.  The mill's position is such that the $x$ and $y$ coordinates of the hub lie at the center of one of the computational cells.

To initialize the simulations, a warm-up simulation is run first, without mill. For this warmup simulation, an inflow-outflow condition is used in the $x$ axis (the main direction of the wind): particles exiting the domain at the outflow boundary are reinserted at the inflow boundary with a mean velocity given by the targeted log-law\footnote{ $u(z) = \frac{\ustar}{\kappa} \log\, \frac{z}{z_0}$.} plus a  random velocity vector normally distributed, that renders the covariance structure taken from the cell where the particle was in the previous time step. A similar boundary condition is used in the $y$ axis.  The results of the warmup simulation are averaged along the $x$ and $y$ axes to produce empirical profiles of all relevant statistics of the velocity field.

For the mill simulations, an inflow-outflow boundary condition is applied in the $x$ axis, using the empirical profile from the warmup simulation to force the inflow condition: particles exiting the domain at the outflow boundary are reinserted at the inflow one with velocities following a trivariate Gaussian distribution computed with the information of these empirical profiles. A similar strategy is used in the $y$ axis boundaries: particles exiting the domain through one of these boundaries are reinserted in the opposite boundary, with a new velocity following a trivariate Gaussian distribution computed with the empirical profiles. 

In all simulations, the boundary conditions for the $z$ axis are as described in Section \ref{sec:sdm}. 

As in \cite{wu-porteagel_2011}, we use the lift and drag data $C_L(\alpha)$ and $C_D(\alpha)$ provided by Blade data  taken from Sunada  Sakaguchi and Kawachi \cite{sunada-etal-1997}.  The chord length and twist angle of the turbine are taken from \cite{wu-porteagel_2011}. 

\subsection{Comparison with experimental data}

The results of SDM with the  
more complex Rotating Actuator Disc mill model is compared against experimental data obtained from  Wu and Port\'e-Agel \cite{wu-porteagel_2011}. Three quantities are of particular interest: the streamwise component of the mean velocity field, the streamwise  turbulent intensity, and the shear stress  between streamwise and vertical components.  As underlined in \cite{wu-porteagel_2011}, du to the non-uniform (logarithmic) mean velocity profile of the incomming boundary-layer flow, the profiles of those three quantities yield  non-axisymmetric distribution. 

SDM simulations are showed without any time average on the SDM output produced after a total of  1000 time iterations after the warm-up phase.  We show $\av{u}$, $I$ and $-\av{uw}$ as they are computed at the final time step, using the second order CIC-estimator for the computation of Eulerian fields on the  last 20 time steps only,  to not burden the  computation time.

\begin{figure}[h!]
\centering
\includegraphics[scale = 0.45]{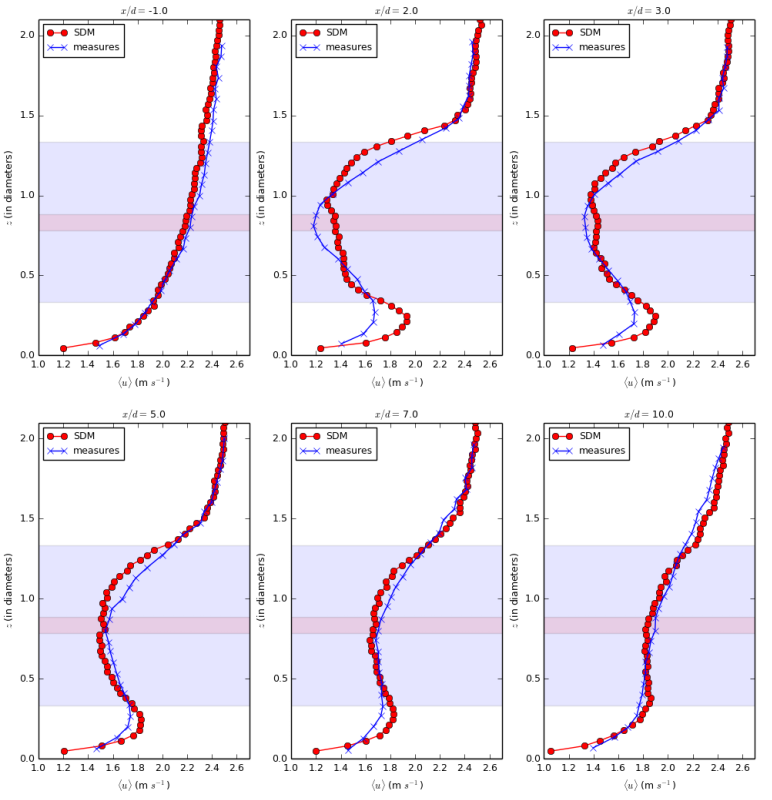}
\caption{Comparison of vertical profiles of $\av{u}$ at different downstream positions $x$ from the turbine (-1, 2, 5, 7 and 10 diameters respectively). The profiles are centred to the hub $y$ position. The blue curve represents the wind tunnel measures, the red curve represents SDM simulation with the Rotating Actuator Disc mill model.} \label{fig:MeanProfiles-tunel} 
\end{figure}

First, we examine the streamwise component of the mean velocity.  A profile comparison against experimental data at several downstream locations is shown in Figure \ref{fig:MeanProfiles-tunel}. For the Rotating ADM a good agreement is seen for distances equal and higher than 5 turbine diameters, while a reasonable agreement is seen near the turbine; 

\begin{figure}[h!]
\centering
\includegraphics[scale = 0.45]{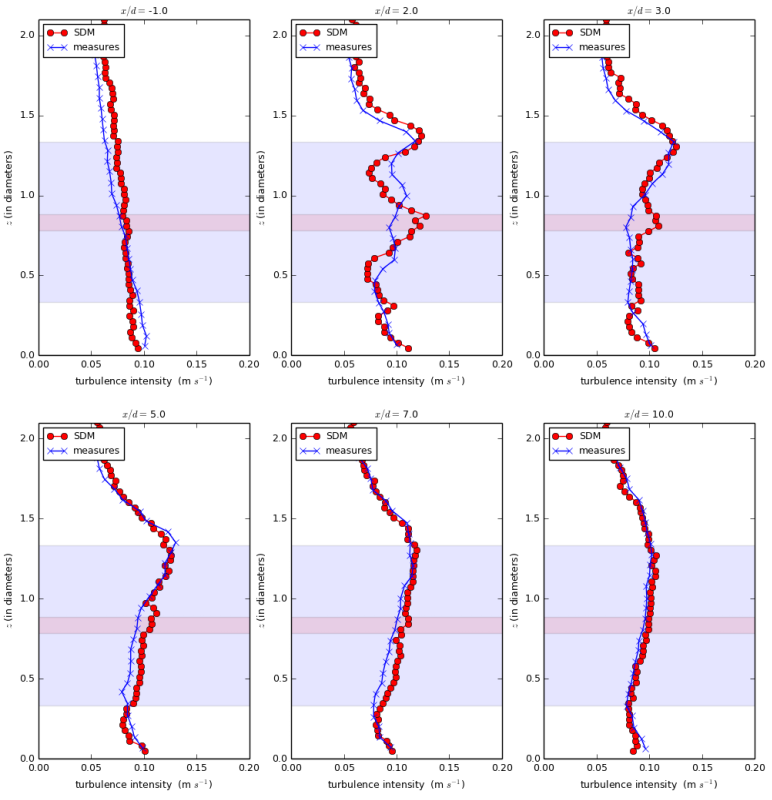}
\caption{Comparison of vertical profiles of the turbulence intensity at different downstream positions $x$ from the turbine (-1, 2, 5, 7 and 10 diameters respectively). The profiles are centred to the hub $y$ position. The blue curve represents the wind tunnel measures, the red curve represents SDM simulation with the Rotating Actuator Disc mill model.}\label{fig:turbulenceIntensityProfilesTunnel}
\end{figure}

Secondly, the turbulence intensity $I$ of the stream-wise component of velocity  is analyzed. 
Figure~\ref{fig:turbulenceIntensityProfilesTunnel} shows the vertical turbulence intensity  profiles, which are plotted at the same downstream positions as was done for $\av{u}$.    
We compute the  turbulence intensity with SDM,  using the same inflow mean velocity $U_\textrm{hub}$ at the hub height than in  \cite{wu-porteagel_2011}:
$$
I  = \frac{\sqrt{\frac{2}{3}\, \tke} }{U_\textrm{hub}},\quad \mbox{ with } U_\textrm{hub} = 2.2 \mbox{ m s}^{-1}. 
$$ 
We observe a good fit with the measures far and close to the mill, even if the turbulent intensity seems to be overestimated in the area of the nacelle (in light red) at 2 and 3 diameters of the hub.   

\begin{figure}[h!]
\centering
\includegraphics[scale = 0.45]{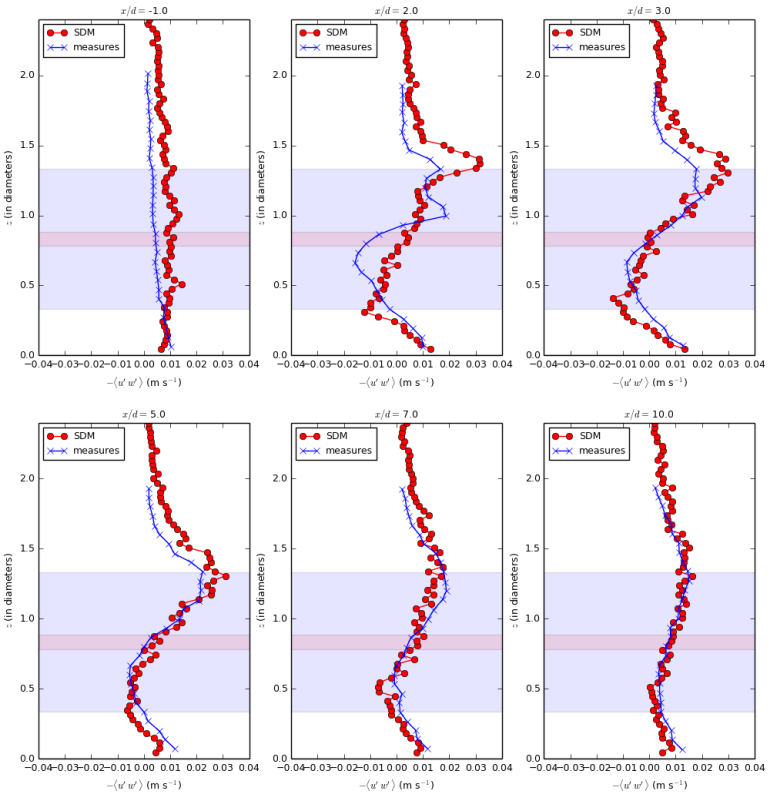}
\caption{Comparison of vertical profiles of the shear stress $-\av{u'w'}$ at different downstream positions $x$ from the turbine (-1, 2, 5, 7 and 10 diameters respectively). The profiles are centred to the hub $y$ position. The blue curve represents the wind tunnel measures, the red curve represents SDM simulation with the Rotating Actuator Disc mill model. }\label{fig:CovarianceProfilesTunnel}
\end{figure}

Also of interest is the shear stress (i.e $-\av{uw}$ the covariance between $x$ and $z$ components of velocity). As before,  vertical profiles are plotted for different downstream positions from the turbine. Figure~\ref{fig:CovarianceProfilesTunnel} shows the results. Again, the fit is good but SDM seems to overestimate a little the  shear stress.  It is  it is worth noting that the SDM computation locates the maximum of the shear stress at the top tip of the blades area (top limit of the light blue area) as expected. The same effect can be observed for the turbulent intensity.  

\begin{figure}[h!]
\centering
\includegraphics[width=0.39\textwidth]{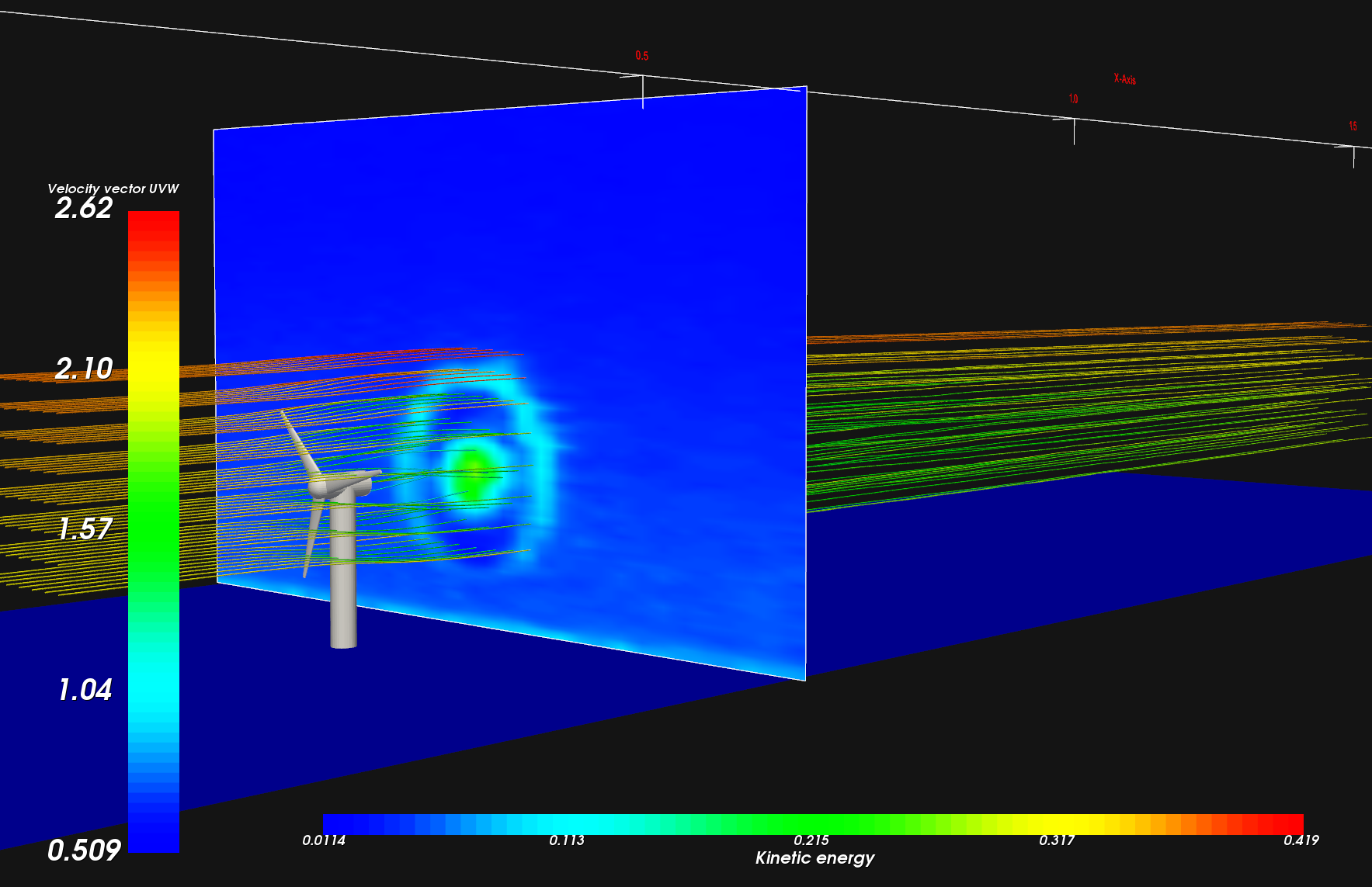}
\includegraphics[width=0.45\textwidth]{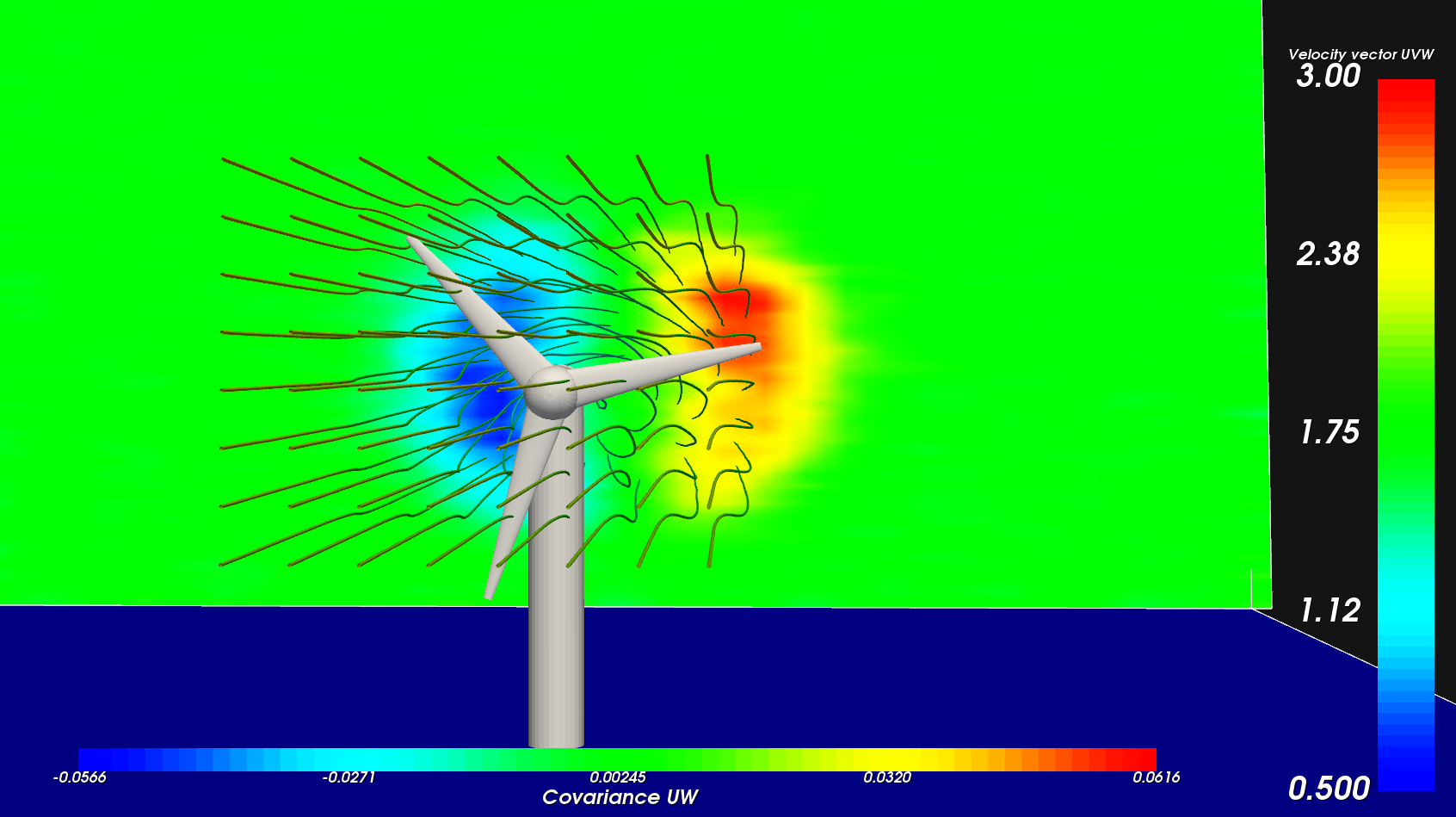}
\caption{Streamlines  visualization (Rotating ADM simulation),
with a  $yz$-contour plot of the turbulent kinetic energy circles at one diameters of the hub (left); with a $yz$-contour plot of the $\av{u'w'}$ covariance at 3.5  diameters of the hub (right).} \label{fig:WindTunnelGUI} 
\end{figure}

We complement the comparison between the 2D  $xz$-profiles in Figures \ref{fig:MeanProfiles-tunel}  \ref{fig:turbulenceIntensityProfilesTunnel} \ref{fig:CovarianceProfilesTunnel},  with some other directional views in Figure \ref{fig:WindTunnelGUI} showing in particular the winding of the streamlines  passing through the turbine. 

\medskip
It is worth to notice that in this first Lagrangian approach for Rotating-ADM, we did not dissociate the $\Delta x$ corresponding to the thickness of the $C_\blades$ cylinder with the $\Delta x$ of the cell mesh. Thus, in these simulations the $C_\blades$ thickness is about 66\% of the disk diameter. Refine this thickness independently to the cell mesh size can bring substantial improvements without additional computational cost.

Furthermore, the model of permeable disk for the nacelle can be improved (without counting the mast model that we have not put yet in our simulation). In these simulations, we introduce the corresponding force term in the velocity computation without any correction terms for the second order moments, as we did for the wall law model at the ground. 
There is therefore some margin of improvement for the Rotating-ADM model with SDM.

\paragraph{Computation time.}  To produce all the results shown in this paper, we have used one  {\it 32-cores Intel Xeon CPU E5-2665 0 @ 2.40GHz} computer node.  The elapsed time for the 1000 iterations with $96\times 33\times 84\times 150$ Lagrangian particles (about 40 millions of particles) is about 8 hours and 10 minutes. A new version of our code for multi-nodes computer architecture is under development. 

\section{Some numerical experiments at the atmospheric scale}\label{sec:simulations}

\begin{figure}[h!]
\centering
\includegraphics[width=0.9\textwidth]{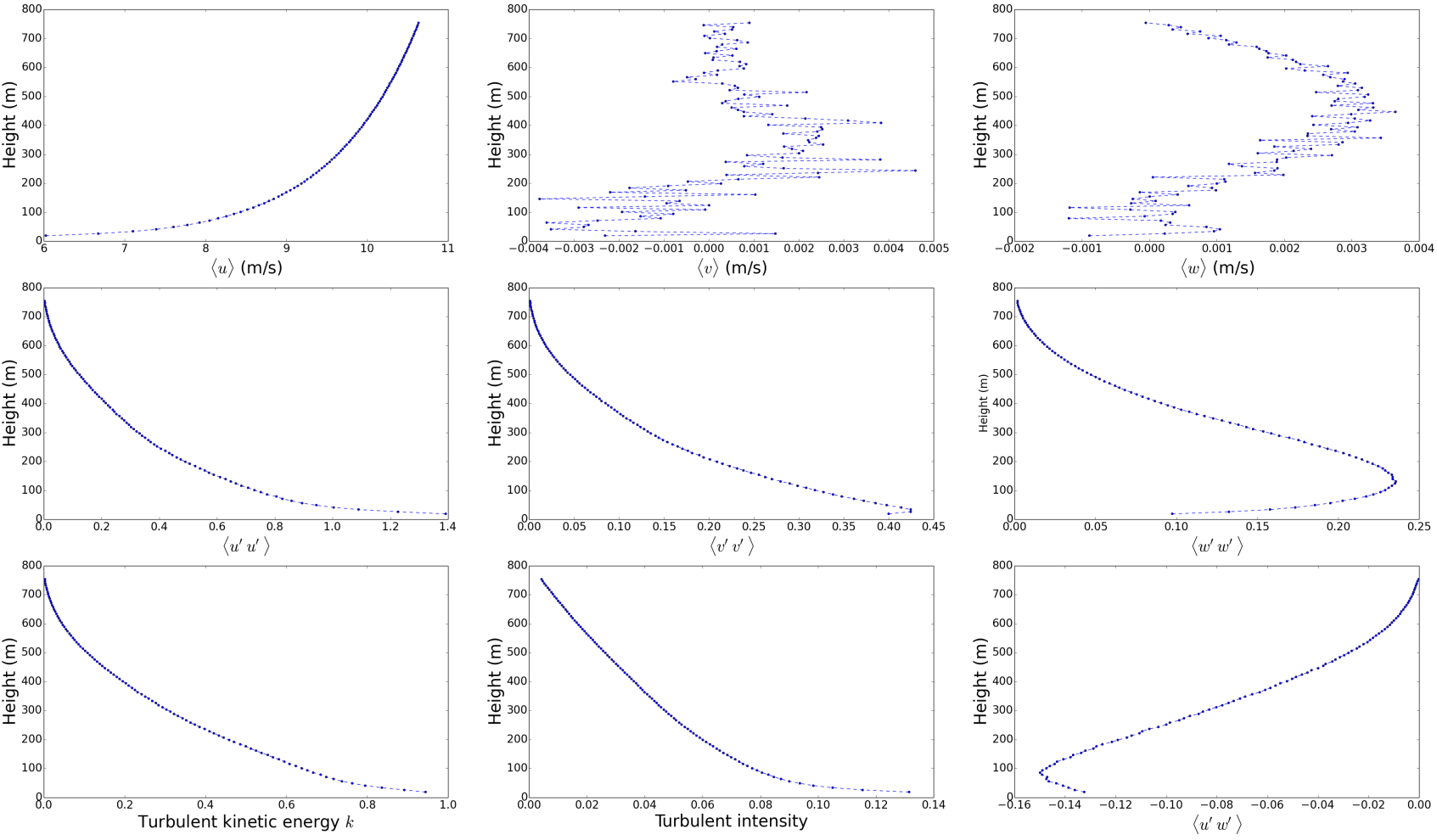}
\caption{Profiles of main turbulence charateristics computed by SDM in the whole boundary layer (averaged in the $x$ and $y$-directions).  The numerical parameters are the same than in Tables \ref{tab:simudata}  exept  that the computational domain height is 750 m.} \label{fig:atmosBL_main_stat}
\end{figure}

We reproduce atmospheric turbulent condition approaching  a real-size neutrally-stratified boundary layer condition,  by parametrizing  our simulation inspired by the numerical experiments performed in Drobinski et al.  in \cite{drobinski-etal_2007} for an atmospheric neutral case. 
The corresponding boundary layer characteristics are summarized in Table \ref{tab:atmospheric_PhC}
Figure \ref{fig:atmosBL_main_stat}  shows how SDM reproduces turbulence characteristics of the same order  than in \cite{drobinski-etal_2007} as well as  log-law profile for the velocity. In particular the anisotropic variance profiles, computed here in the whole boundary layer depth converge to zero at the top of the boundary layer. The turbulent kinetic energy goes naturally to zero at the boundary layer top and SDM fits the prescribed laminar geostrophic flow. 

\subsection{Simulation setup}

\begin{figure}[ht]
	\centering
	\includegraphics[width=0.6\textwidth]{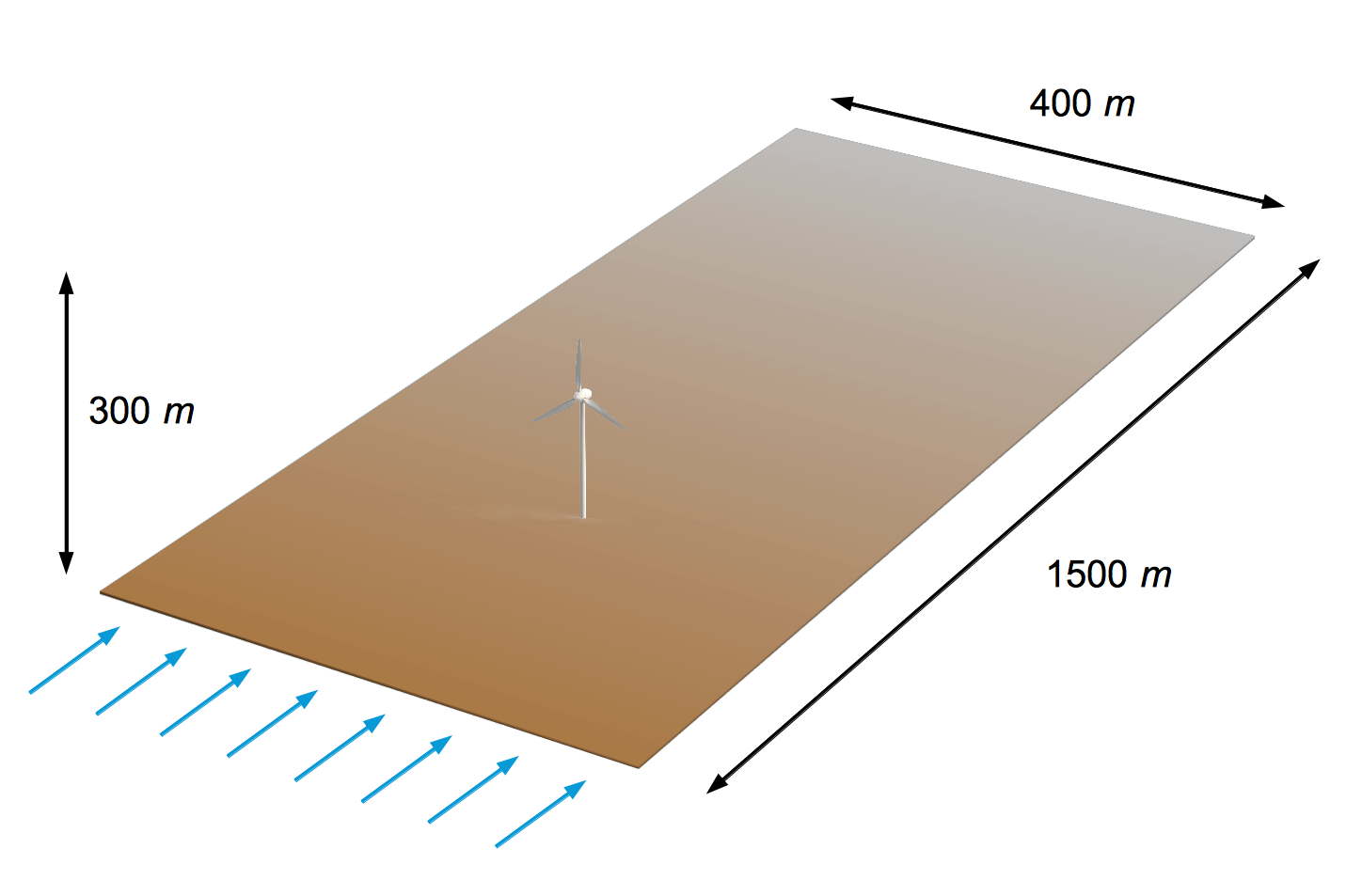}
	\caption{Domain for the atmospheric scale simulations. \label{fig:domain}}
\end{figure}

To test the implementation of mills in the stochastic Lagrangian settings presented before, a one-mill configuration has been used. A single mill has been placed in a rectangular domain spanning $1500\times 400\times 300$~meters in the $x$, $y$ and $z$ directions, respectively (Figure~\ref{fig:domain}). The mill faces atmospheric flow with a log-law profile at the inlet section, that develops moving along the $x$ direction. The main physical and computational parameters of the simulations are detailed in Table~\ref{tab:simudata}.

\begin{table}[h!]
\centering
\subfloat[Main characteristics of the boundary layer flow \label{tab:atmospheric_BL}]{
\begin{tabular}{| l | l |}
	\hline
	\multicolumn{2}{|c|}{\textbf{Boundary layer characteristics}} \\
	\hline
	Boundary layer depth  & 750 m \\
	\hline
	Wind speed at the top & 10.63 m\,$\text{s}^{-1}$ \\
	\hline
    Velocity friction $\ustar$ & 0.42 m\,$\text{s}^{-1}$ \\
	\hline
	Roughness length $z_0$ & 0.03  m \\
	\hline	
	\end{tabular}
}\quad
\subfloat[{Main model physical constants\label{tab:atmospheric_PhC}}]{
\begin{tabular}{| l | l |}
	\hline
	\multicolumn{2}{|c|}{\textbf{Model constants}}\\
	\hline
	Rotta constant $C_R$  &  1.8\\
	\hline
	von Karman constant $\kappa$ & 0.4 \\
	\hline
    $C_2$  &  0.60\\
	\hline
	$C_\epsilon$ & 0.08\\
	\hline
	$\zl$ & 150 m\\
	\hline
	\end{tabular}
}
\\
\subfloat[configuration of the simulations]{
\begin{tabular}{| l | l |}
\hline
\multicolumn{2}{|c|} {\textbf{Simulation parameters}} \\
\hline
Domain size $x$ & 1488 m  \\
\hline
Domain size $y$ & 403 m \\
\hline
Domain size $z$ & 300 m \\
\hline
90 cells in $x$ & $\Delta x =$ 16  m  \\
\hline
31 cells in $y$ & $\Delta y =$13 m\\
\hline
80 cells in $z$ & $\Delta z =$ 3.75 m\\
\hline
Particles per cell & 128 \\
\hline
Final time is 1000 s & Time step is 1.0 s \\
\hline
\end{tabular}}
\quad 
\subfloat[parametrers of the mill]{
\begin{tabular}{| l | l |}
\hline
\multicolumn{2}{|c|} {\textbf{Mill configuration}} \\
\hline
Coordinates of the hub: & (496.875,200,50) m \\ 
\hline
Hub height & 50 m  \\
\hline
Radius & 20.5 m \\
\hline
Nacelle radius & 4.5 m \\
\hline
Rotational speed & 2.83 $\text{rad.s}^{-1}$ \\
\hline
$a_\nacelle$ for the  Rotation ADM & 0.38 \\ 
\hline
$a_\nacelle$ for the  Non Rotation ADM  & 0.45 \\
\hline\end{tabular}}
\caption{Main parameters of the simulations at the atmospheric scale. \label{tab:simudata}}
\end{table}

To assess the impact of the mill in the flow, simulations are run with  the two turbine models presented in Section \ref{sec:actuator}, namely: the \textit{non-rotating actuator disc} mode (NR-ADM), and the \textit{rotating actuator disc} model (R-ADM).  The mill's position is such that the $x$ and $y$ coordinates of the hub lie at the center of one of the computational cells.

To initialize the simulations,  a warm-up simulation is run  first, according to the same process described to initialize the wind tunnel simulation in Section \ref{sec:validation}. 	 

For the mill simulations, an inflow-outflow boundary condition is applied in the $x$ axis,  as described again in  Section \ref{sec:validation}. 	 
    
For the rotating case, a realistic wind turbine has been designed using blade data of a NTK 500/41 wind turbine (as found in \cite{hansen_2008}), together with lift and drag coefficients corresponding to a NACA 23012 airfoil at Reynolds numbers of the order of $10^6$ (as found in  Abbott and von Doenhoff~\cite{abbott_1959}).

\subsection{Computation of the non-rotating equivalent}
For the non-rotating case, we need to specify the values of $a$, $C_{T}$ and  $\uinf$ (as used in Equations \eqref{eq:inflowFactor}, \eqref{eq:thrustCoeff}, and  \eqref{eq:force2}) so that the simulation is equivalent to the rotating actuator disc. Again, we define this equivalence by requiring the total thrust force in both cases to be the same. However, in this section we introduce an additional method to compute the required quantities, which also serves as a consistency validation of our simulations in the atmospheric case: the Blade Element Momentum method (BEM).

\subsubsection*{Compute thrust from the SDM particles}

For the non-rotating (NR) actuator disc model, we compute the total thrust excluding the area $A_{\nacelle}$ occupied by the nacelle: 
\begin{equation}\label{eq:thrust1}
F_{x}^{\text{NR}} = -\frac{1}{2} \rho (A - A_{\nacelle}) C_{T} \uinf^2, 
\end{equation}
whereas for the rotating (R) model, the total thrust can be computed by integrating Equation \eqref{eq:axialForceTotal.and.tangentialForceTotal} from the nacelle radius $r_{\nacelle}$ to the turbine radius $R$:
\begin{equation}\label{eq:thrust2} 
F_{x}^{\text{R}} = -N_\blades \int\limits_{r_{\nacelle}}^{R} \left( \frac{dL}{dr}\cos\left(\phi\right) + \frac{dD}{dr}\sin\left(\phi\right) \right) dr.
\end{equation}

The idea is then to compute $\uinf$ and $F^{\text{R}}_{x}$, and substitute them in \eqref{eq:thrust1} to obtain $C_{T}$. The magnitude $\uinf$ can be easily estimated as the $z$-averaged value of the mean velocity $\av{\Uu}$ at the inlet section of the domain, over the diameter of the turbine:
\begin{equation}\label{eq:uInfty} 
\uinf = \frac{1}{2R}\int\limits_{h-R}^{h+R} \frac{\ustar}{\kappa} \log\left(\frac{z}{z_{0}}\right) dz, 
\end{equation}

To compute $F_{x}^{\text{R}}$ in  \eqref{eq:thrust2} from the particle information, the integrand is estimated for all particles within region $\mathcal{C}$, using equations \eqref{eq:axialForce4.and.tangentialForce4}: 

\begin{equation}\label{eq:integrand}
\frac{1}{\rho}\dfrac{dF^{\text{R}}_{x}}{ dr} = \frac{1}{\rho} \left( \frac{dL}{dr}\cos\left(\phi\right) + \frac{dD}{dr}\sin\left(\phi\right) \right) \simeq  -2 \pi r\left(\Xx_{t}\right) \Delta x f_{x} \left(\Xx_{t},\Uu_{t}\right). 
\end{equation}

The total thrust over $\rho$, $F_{x}^{\text{R}} / \rho$, is then computing by an estimation of the integral:

\begin{equation}\label{eq:integral} 
\frac{F_{x}^{\text{R}}}{\rho} = R \;\frac{\displaystyle \sum_{p=1}^{N_p}2 \pi\, r\left(\Xx^{p}_{t}\right) \Delta x f_{x} \left(\Xx^{p}_{t},\Uu^{p}_{t}\right) \ind_{\mathcal{C}}\left(\Xx^{p}_{t}\right)  }{\displaystyle \sum_{p=1}^{N_p}  \ind_{\mathcal{C}}\left(\Xx^{p}_{t}\right) } .
\end{equation}

\subsubsection*{Compute thrust with BEM}

A complete description of BEM theory can be found in \cite{hansen_2008} or \cite{manwell-etal_2002}. It stems from the combination of two different analyses of the turbine performance facing a steady-state, uniform, radially symmetric flow: 
\begin{itemize}
\item[(1)] a linear and angular momentum balance in thin, radially distributed, annular stream tubes passing through the turbine swept area at different radii;
\item[(2)] a blade element analysis of the turbine. 
\end{itemize}
The first of these analyses assumes that the relative velocity decrease from the far upstream region to the disc region depends on the radial position, so that $a$ now varies with $r$:
\begin{equation}\label{eq:inflowFactor2} 
a(r) = \frac{\uinf - U_{D}(r)}{\uinf}.
\end{equation}

It is also assumed that the flow gains angular momentum, related to the rotational speed of the turbine. The angular speed $U_{\theta}$ at radius $r$ at the disc is controlled by an additional function $a^{\prime}(r)$:

\begin{equation}\label{eq:tangentialInflowFactor} 
U_{\theta}(r) = a^{\prime}(r) \omega r, 
\end{equation}
where $\omega$ is the angular speed of the turbine. The following expressions are then found for the differential forces in directions ${\bf{e}}_{x}$ and ${\bf{e}}_{\theta}$, at an annulus located at radius $r$ from the turbine's center (see \cite{manwell-etal_2002} or \cite{hansen_2008} for details):

\begin{align}
dF_{x}(r) &= -4\pi \rho r a(r)(1-a(r)) \uinf^2 dr, \label{eq:diffAxial1}  \\
dF_{\theta}(r) &  = 4\pi \rho \omega a^{\prime}(r)(1-a(r)) \uinf r^2 dr. \label{eq:diffTangential1} 
\end{align}
 
Given the assumption of radially symmetric flow made by BEM, in this analysis $\Urel$ and the angle $\phi$ are functions of $r$ only. In terms of $a(r)$ and $a^{\prime}(r)$, expressions \eqref{eq:relativeVelocity} and \eqref{eq:flowAngle} are simplified to:
\begin{equation}\label{eq:relVelocity2} 
\Urel(r) = (1-a(r))\uinf{\bf{e}}_{x} + (1+a^{\prime}(r)) \omega r {\bf{e}}_{\theta},
\end{equation} 
\begin{equation}\label{eq:flowAngle2} 
\phi(r) = \arctan\left(
\frac{(1-a(r)) \uinf}{ (1+a^{\prime}(r)) \omega r }
\right).
\end{equation}
On the other hand, a pure blade element analysis like the one presented in Section \ref{subsec:ADR} yields the following expressions for the same quantities:
\begin{align}
dF_{x}(r) & = -\frac{1}{2}N_\blades c \rho\; \urel^2(r) ( C_{L}\cos (\phi(r)) + C_{D}\sin(\phi(r)) ) dr, \label{eq:diffAxial2}  \\
dF_{\theta}(r) & = \frac{1}{2}N_\blades c \rho\; \urel^2(r) ( C_{L}\sin (\phi(r)) - C_{D}\cos(\phi(r)) ) dr.\label{eq:diffTangential2} 
\end{align}

In BEM, the rotor is discretized in a finite number of blade elements. For a set of radial positions and given values of $\uinf$ and $\omega$, the values of $a$ and $a^{\prime}$, the flow angle $\phi$ and the relative velocity magnitude $\urel$ are computed for each blade element separately, by equating \eqref{eq:diffAxial1} with \eqref{eq:diffAxial2} and \eqref{eq:diffTangential1} with \eqref{eq:diffTangential2} and using an iterative procedure (see \cite{hansen_2008} for a detailed description of the algorithm). The procedure delivers the values of $\urel(r)$, $\phi(r)$, $a(r)$ and $a^{\prime}(r)$.

\paragraph{Two Non-Rotating equivalent models are obtained:} one with the computation of the thrust according to \eqref{eq:integral}, and the other using the BEM method described above. 
The corresponding estimated values for the thrust coefficient  are reported in Table ~\ref{table:thrust}. The $a$ values produced par the two methods are so close that the plotting of the corresponding simulation profiles (as those  in Figures \ref{fig:meansU} and \ref{fig:variancesU}) are completely  indistinguishable. This first comparison validates the thrust equivalence computation with the SDM particles, but as we will see in the next section, at least with this value of $a$, the NR-ADM  underestimates   the  three main characteristic that we analyze in the flow, $\av{u}$, $\av{u'u'}$ and $\av{u'w'}$.  

In addition to the assumptions of homogeneity and symmetry of the wind fields that are not well respected in our case, the thrust computation methods described above may be  sensitive to the blade geometry discretization parameter (in our data $dr =$1m) for the BEM-based method,  to the $\Delta x$ parameter for the density estimator, in \eqref{eq:integral} for the SDM particles-based method. 

\begin{table}[h!]
\centering
\begin{tabular}{|c|c|c|c|}
\hline
\multicolumn{4}{|c|} { \textbf{Equivalent non rotating actuator disc parameter estimations}} \\[0.1cm]
\hline
Method & $\ds F^{\text{R}}_{x}/\rho$ ~($N_\blades \text{m}^{3}/\text{kg}$) & ${a}$ & $C_{T}$ \\[0.1cm]
\hline 
SDM Particle computation & 
{28680.55} & 
{0.2081584} & 
{0.659314}
\\[0.1cm]
\hline
BEM computation & {27863.06} & {0.2034322} & 
{0.648190} 
\\[0.1cm]
\hline
\end{tabular}
\caption{Estimation of the equivalent non-rotating actuator disc for our mill configuration.  \label{table:thrust}}
\end{table}

\subsection{Numerical experiments, comparison between rotating and non-rotating actuator disc methods}

It is interesting and useful to analyze the differences between results obtained with the simple Non-Rotating model and the more complex Rotating model, and to determine to which extent they give a similar development of the turbine wake. In particular, as can be seen in the simulations of Wu and Port{\'e}-Agel \cite{wu-porteagel_2011}, one would expect an under-prediction of the speed deficit near the turbine for the non-rotating case, while both models should yield similar results far downstream from the turbine. This and other effects are studied hereafter. 

\begin{figure}[h!]
\centering
\includegraphics[width=9cm, height=3cm]{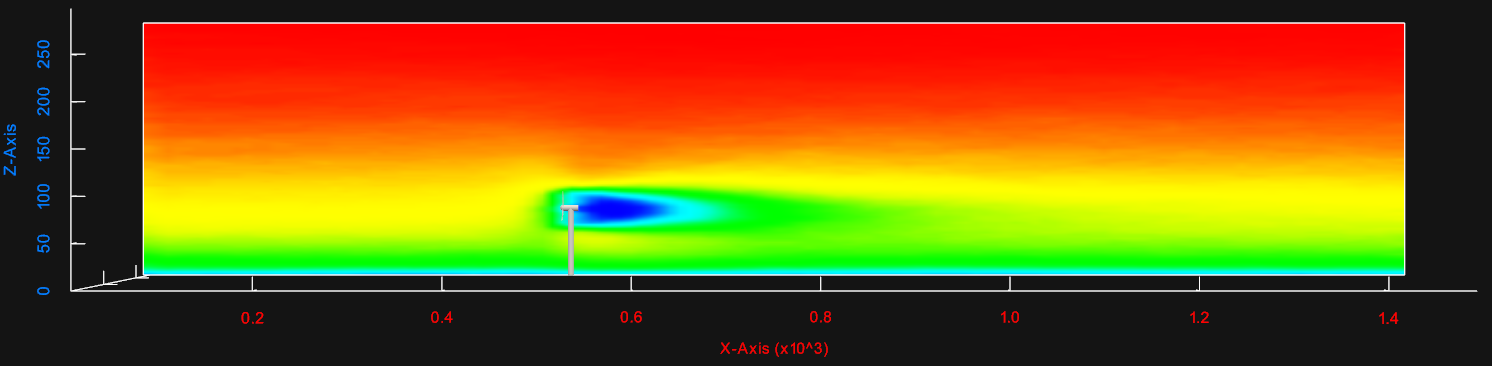} 
\includegraphics[height=3cm]{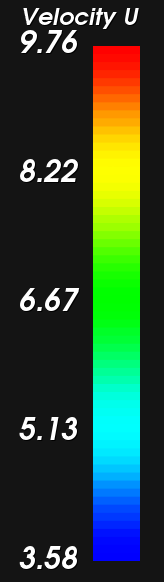} 
\includegraphics[width=9cm, height=3cm]{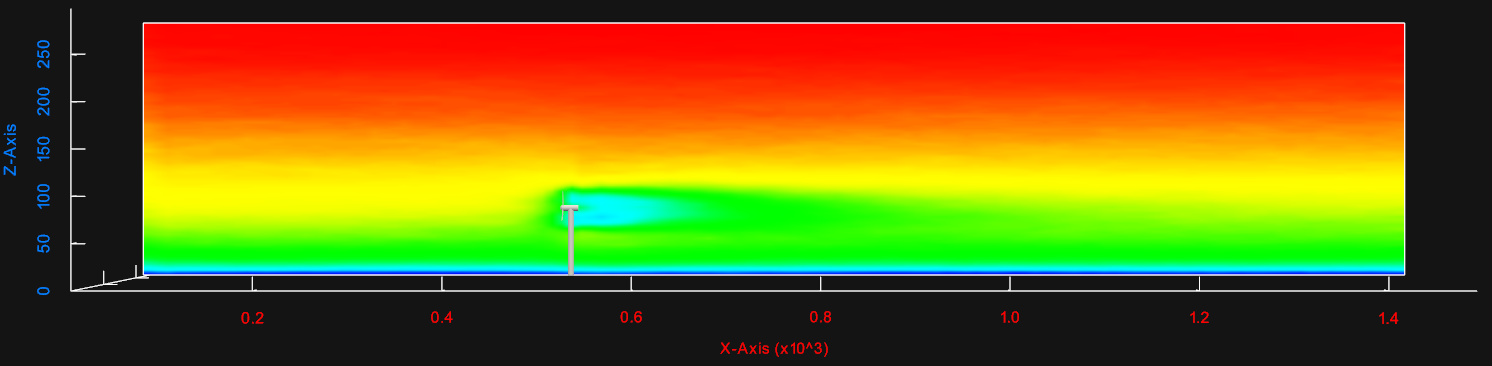} 
\includegraphics[height=3cm]{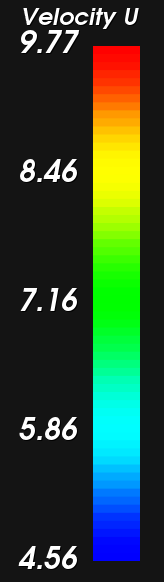} 
\caption{Comparison of $xz$ cross sections of $\av{u}$: Rotating Actuator Disc (top), Non-Rotating Actuator Disc with $a$ estimated from thrust equivalence (bottom).}\label{fig:meansU}
\end{figure}

\begin{figure}[h!]
\centering
\includegraphics[width=0.5\textwidth]{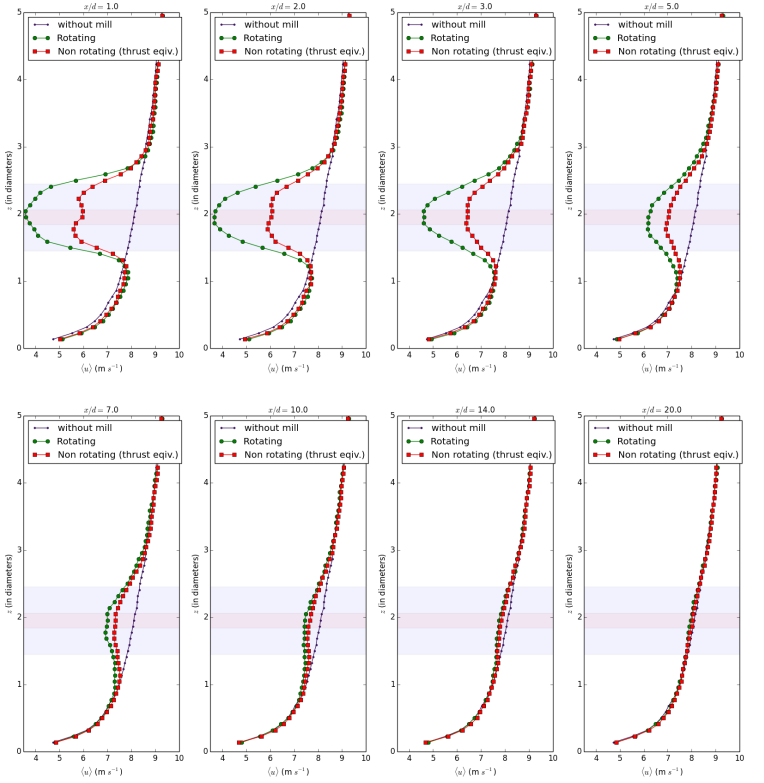}
\caption{$\av{u}$ profiles at different downstream positions from the turbine.} \label{fig:figureMeanProfiles} 
\end{figure}

To begin with, mean velocity contour plots along the stream-wise direction are shown  for both models in  Figure~\ref{fig:meansU}. 
These plots show that  rotation and non-uniform loading have a clear impact in the turbine wake. This is confirmed by the mean velocity profiles at different downstream positions, as shown in Figure~\ref{fig:figureMeanProfiles}. As expected the Rotating AD model and the Non-Rotating AD model present substantial differences, as  already observed in the LES framework in \cite{wu-porteagel_2011}. The mean velocity component $\av{u}$ is underestimates in the disk area (blades area corresponds to the light blue zone, and the nacelle area corresponds to the light red zone  in the plots). 

\begin{figure}[h!]
\centering
\includegraphics[width=9cm, height=3cm]{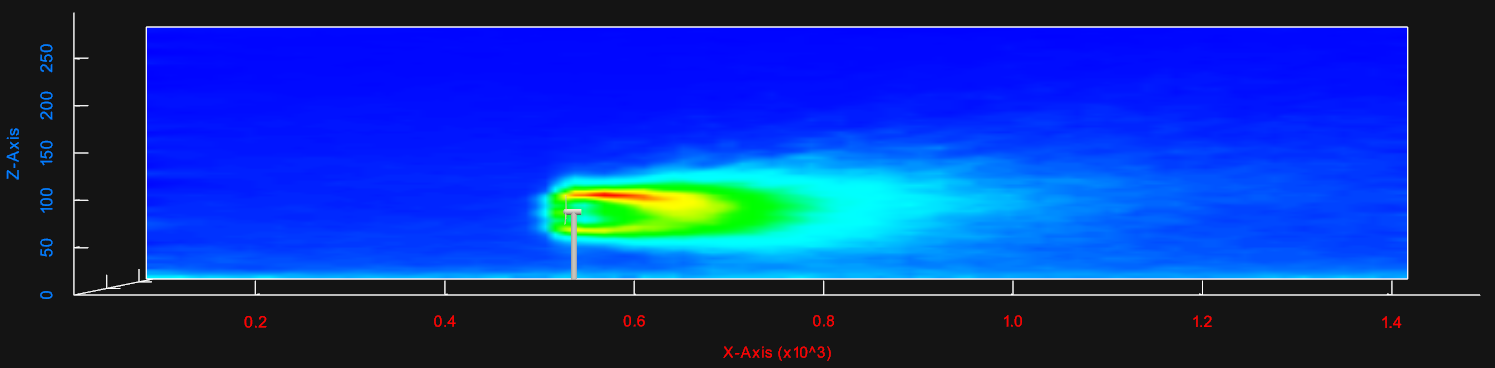} 
\includegraphics[height=3cm]{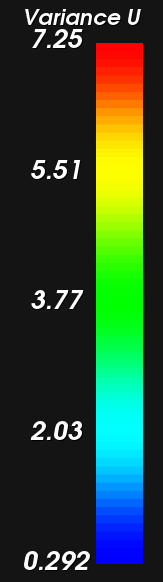} 
\includegraphics[width=9cm, height=3cm]{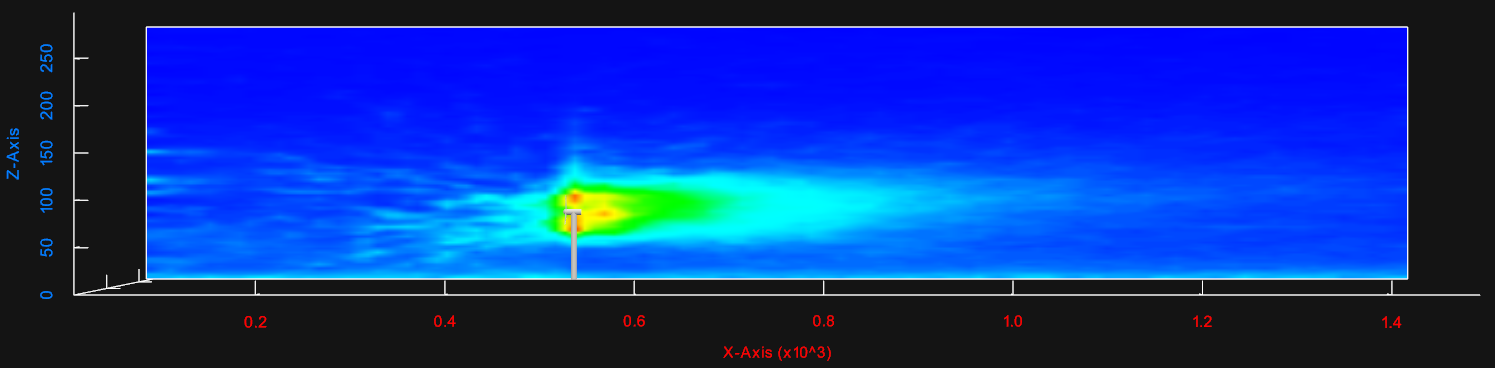} 
\includegraphics[height=3cm]{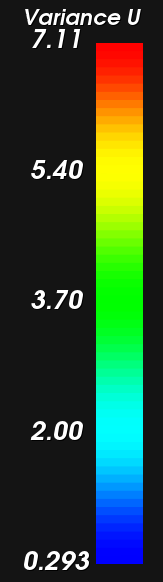} 
\caption{Comparison of $xz$ cross section of variance $\av{u'u'}$ along $x$: Rotating Actuator Ddisc (top), Non-Rotating Actuator Disc with $a$ estimated from thrust equivalence (bottom).}\label{fig:variancesU}
\end{figure}

Secondly, the variance of the stream-wise component of velocity $\av{u'u'}$ is analyzed. Figure~\ref{fig:variancesU} shows contour plots taken at the middle of the $xz$ plane, for both simulations. It is noteworthy that the variance near the turbine is much higher for the rotating actuator disc simulation, although far from the turbine the differences begin to fade. It is also interesting that in Rotating AMD case the highest amount of variance is generated atop the turbine, while for the Non-Rotating AMD the variance at the hub is  uniformly strong along the turbine disc, and not just at the tip top. We can obverse the impact of this variance anomaly that propagates  in the particles environment  behind the mill  to the left and top.
Knowing this, a remedy may consist to impose the values of the second order moments  at the left boundary side from the warmup simulation to overcome  this  weird propagation of the variance. 

\begin{figure}[h!]
\centering
\includegraphics[width=0.5\textwidth]{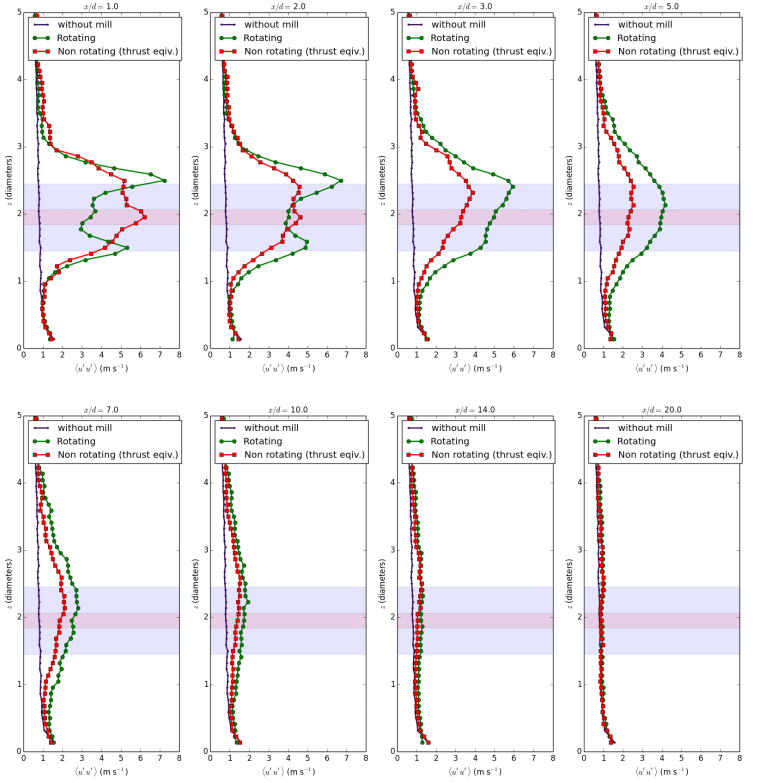}
\caption{$\av{u'u'}$ profiles at different downstream positions from the turbine.}\label{figureVarianceProfiles}
\end{figure}

The same behavior can be seen in the vertical variance profiles, which are plotted at the same downstream positions as was done for $\av{u}$ (see Figure~\ref{figureVarianceProfiles}).  
At one diameter from the turbine, we can observe the overestimation of  $\av{u'u'}$ for the NR-ADM (in red), in particular in the nacelle area. A best estimate of  $a_\nacelle$ could therefore  contribute to reduce this variance anomaly. 

\begin{figure}
\centering
\includegraphics[angle=90,width=6cm, height=4cm]{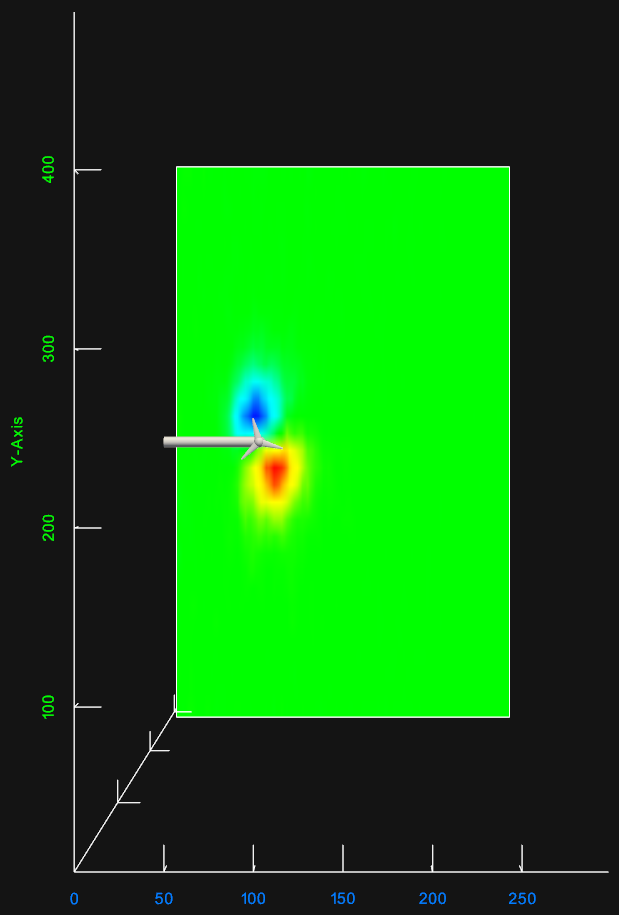} 
\includegraphics[height=4cm]{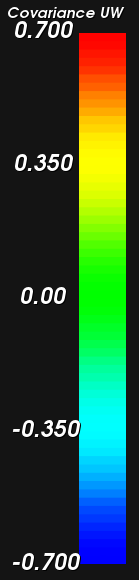} 
\quad
\includegraphics[angle=90,width=6cm, height=4cm]{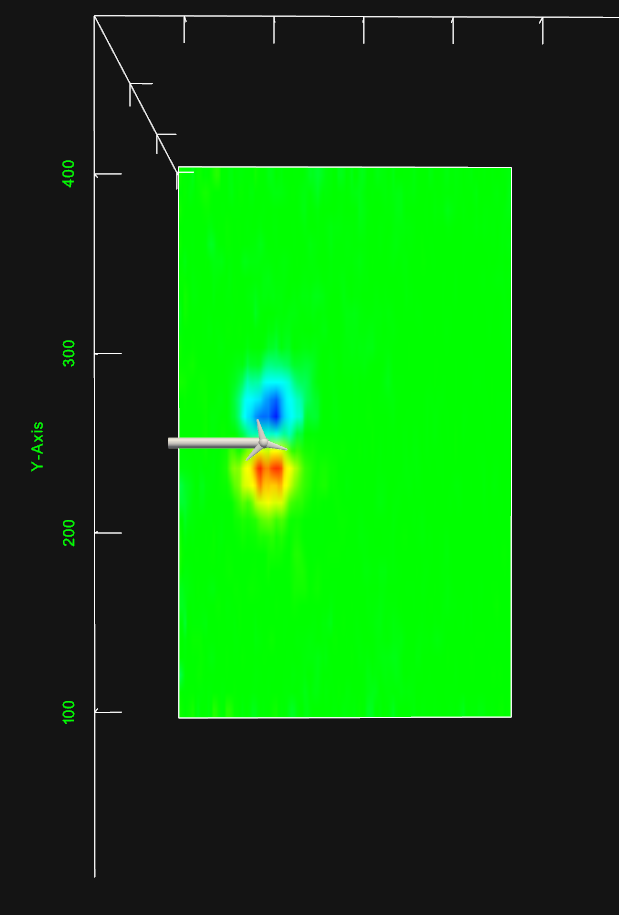} 
\includegraphics[height=4cm]{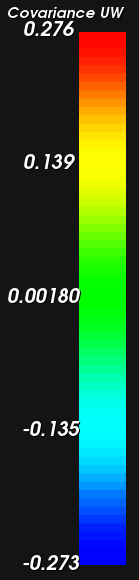} 
\caption{Comparison of the $yz$ cross section of covariance $\av{u'w'}$: Rotating Actuator Disc (left), Non-Rotating Actuator Disc with $a$ estimated from thrust equivalence (right).}\label{fig:covariancesUW}
\end{figure}

Also of interest is the covariance between $x$ and $z$ components of velocity. As before, to compare both simulations, $xz$ contour plots are taken at the middle of the $y$ axis, and vertical profiles are plotted for different downstream positions from the turbine. Again, the rotating model produces stronger values of covariance near the turbine with respect to the non-rotating model, while both tend to equalize as one moves to the far downstream section. Figure~\ref{fig:covariancesUW} shows the results.
\begin{figure}[h!]
\centering
\includegraphics[width=0.5\textwidth]{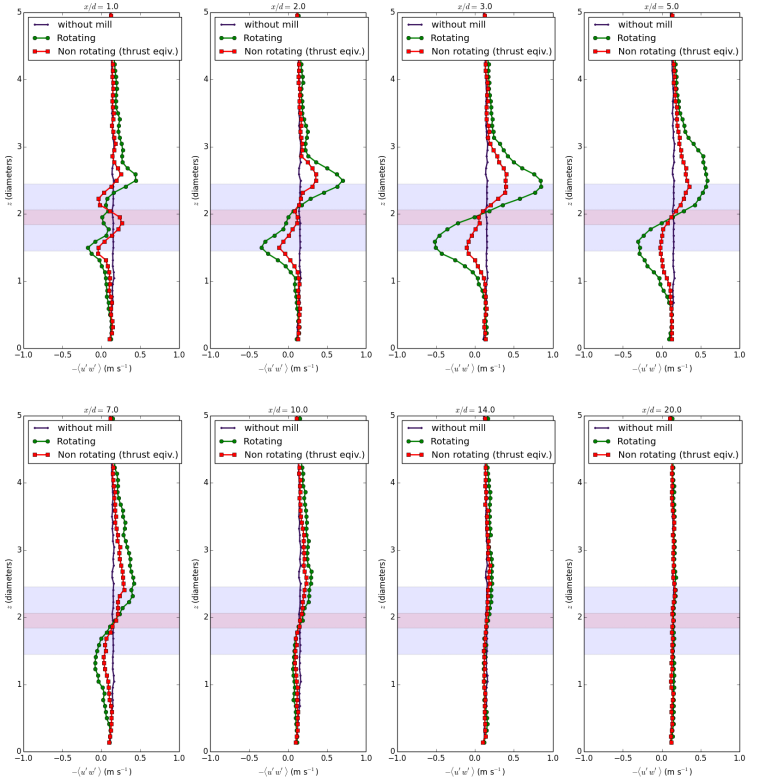}
\caption{$\av{u'w'}$  profiles at different downstream positions from the turbine.}\label{figureCovarianceProfiles}
\end{figure}

Figure~\ref{figureCovarianceProfiles}  gives  some $yz$-cross sections of the covariance at the same $x$-distance to the turbine. The spacial structures of the covariances are similar, but the level of the shear  is about three times smaller for the NR-ADM at 3.5 diameters from the turbine. 

\subsection{Probability distribution functions of the streamwise velocity}

One of the main advantages of the \textit{stochastic downscaling method} presented in Section \ref{sec:sdm} is that it allows access to the instantaneous probability distribution functions of the wind velocity field at each time and position. In the case of mill simulations, this information may be used for various purposes. Contrary to deterministic methods, the stochastic methodology used here estimates the PDFs of the velocity field in just one simulation, directly by sampling the particle properties in the same way the various statistics presented before are estimated. Plus, no time or spatial averaging is required, and it is possible to see how the PDF varies along the wake of a turbine. 

In this section, we present histograms that estimate the PDF of the streamwise velocity component $u = \av{u} + u'$,  corresponding to several points before and after the wind turbine, located at the middle of the $y$ axis and at hub height. To obtain each histogram, we discretise velocity space and sample the particle information of all cells in the neighborhood of the point of interest. 

\begin{figure}[h!]
\centering
\subfloat[-1 diameters from turbine]{\includegraphics[width=.3 \textwidth]{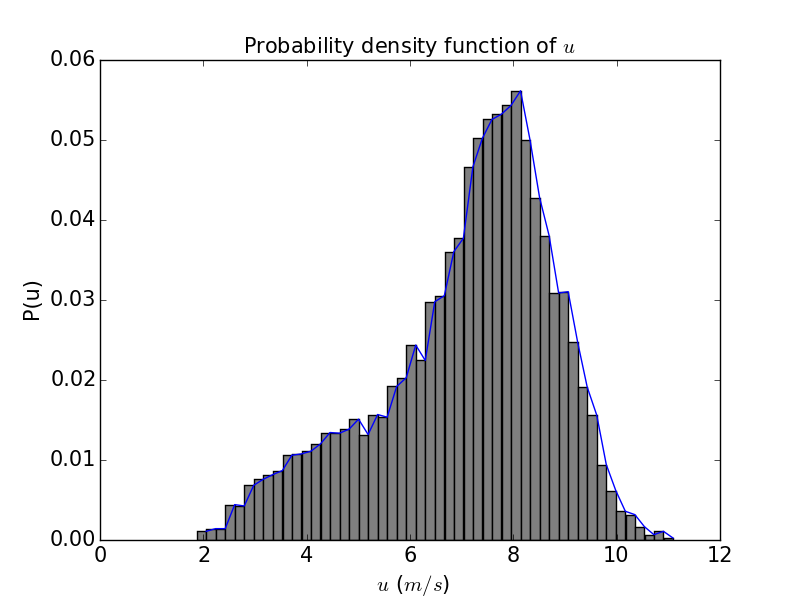}} \quad 
\subfloat[ 2 diameters from turbine ]{\includegraphics[width=.3\textwidth]{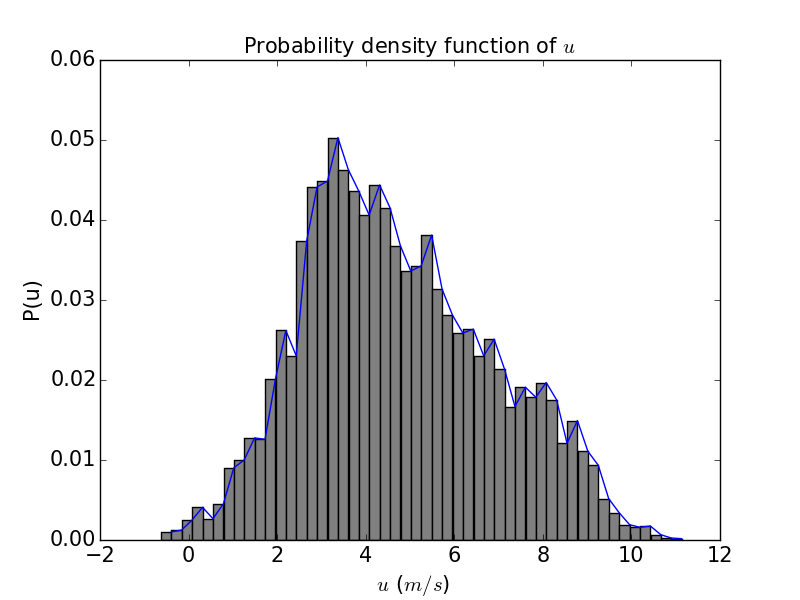}} \quad 
\subfloat[ 3 diameters from turbine]{\includegraphics[width=.3\textwidth]{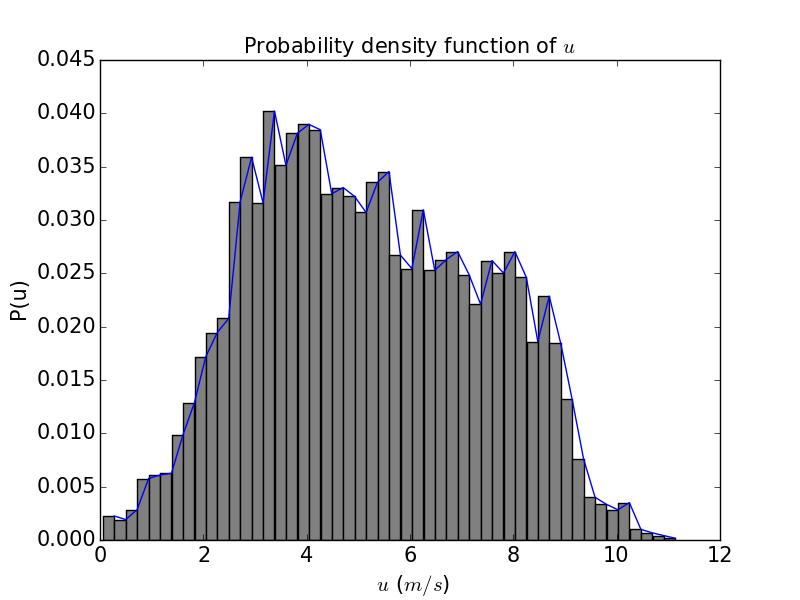}}\\
\subfloat[ 4 diameters from turbine]{\includegraphics[width=.3\textwidth]{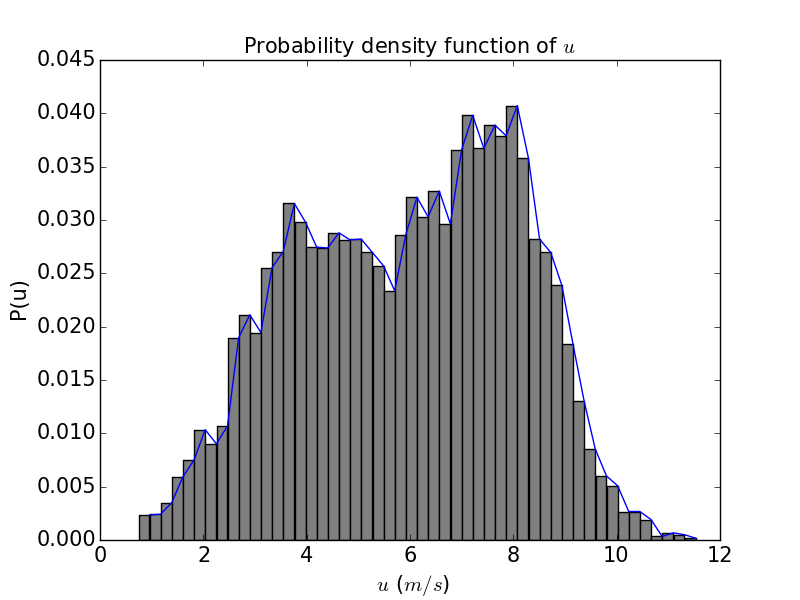}}\quad 
\subfloat[ 5 diameters from turbine]{\includegraphics[width=.3\textwidth]{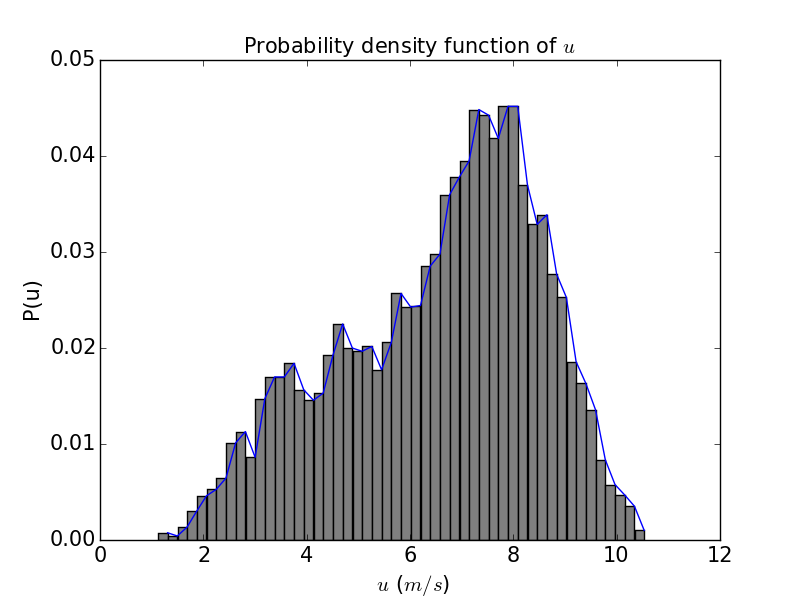}}\quad 
\subfloat[ 7 diameters from turbine]{\includegraphics[width=.3\textwidth]{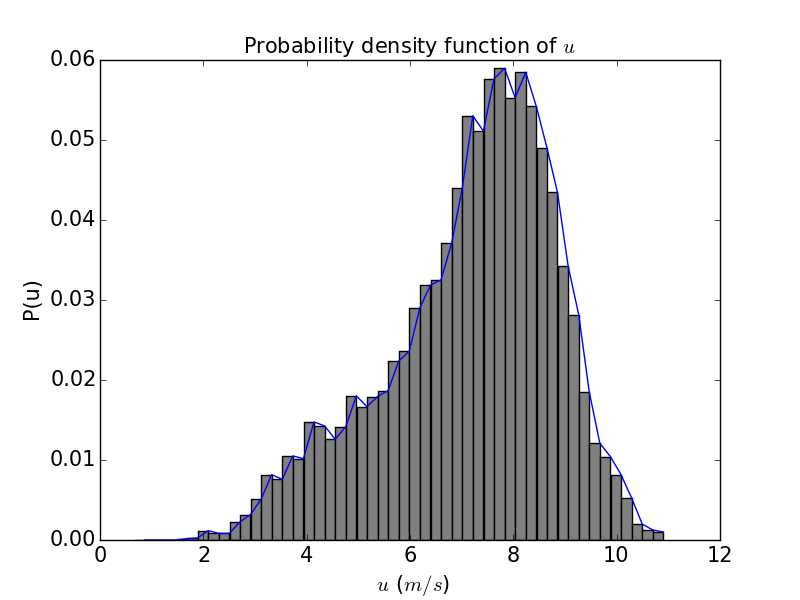}}
\caption{Local distribution of velocity component  $u = \av{u} + u'$ at different positions of the $x$ axis. }\label{figurePDF} 
\end{figure}

As can be seen in Figure~\ref{figurePDF}, the distribution of $u$ is fairly non-symmetric at $-1$ diameter before the turbine (where the flow has not yet felt its presence) with a peak around 8 m s$^{-1}$, 
moving to around 3 m s$^{-1}$ in the near turbine wake, progressively recovering to the $-1$ diameter distribution as one moves downstream from the turbine, and showing a bimodal distribution  transition at 3 and 4 diameters from the turbine.

\section{Conclusions}\label{sec:conclusions}

In recent years, wind energy has seen an important growth worldwide, and the construction and operation of large wind farms necessitates a better understanding of the flow inside and through them in realistic and dynamic atmospheric conditions. In the present study, actuator disc methods have been coupled with the SDM solver \cite{ber-bos-chauv-jabir-rous-09},  which uses a \textit{probability distribution function} (PDF) approach to turbulent flow simulation based on stochastic Lagrangian models. The result is an innovative  methodology to simulate wind turbines and wind farms operating in atmospheric boundary layer flow. It has been shown that the particle setting of SDM is fit for mill simulations, providing qualitatively consistent estimations of the mill forces and wake properties. Plus, the PDF framework coupled with actuator disc methods allows one to obtain an estimation of the wind variability in the wake of a wind turbine, something which - to the best knowledge of the authors - has never been accomplished before. 

The present work is part of a larger project, which seeks to simulate large wind farms with complex topography using the SDM solver. 
Moreover as a downscaling method, SDM aims to be coupled with dynamic boundary conditions coming typically from mesoscale  meteorological solvers. 
Here, the intent has been to present the methodology, as well as to give a preliminary validation analysis with respect to measurements and a comparison of the different models available, using simple attaptation of actuator disc models and just one turbine. Future stages of the project will produce: (1) a wind farm simulation using dynamical downscaled boundary conditions coming from a coarse meteorological solver; 
(2) simulations with complex topography; and (3) implementation of more complex models for the blade forces in the SDM framework. 

\begin{figure}[!ht]
\centering
\includegraphics[scale = 0.55]{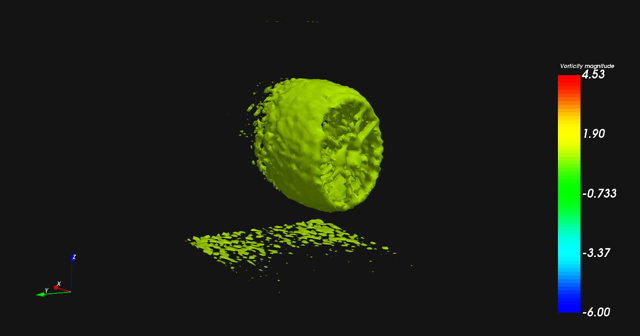}
\caption{Iso-surfaces of the $x$ component of vorticity behind a wind turbine. \label{fig:vorticity}}
\end{figure}

Regarding this last point, we would like to conclude showing some of our ongoing work concerning individual blade visualizations, using the methodology presented in Section \ref{subsec:ADR}. 
In particular, it is possible - by modifying the shape of the blade sectors - to obtain estimations of more involved 3D quantities than the ones presented here, such as the vorticity structures generated behind an individual mill. As proof of concept, Figure \ref{fig:vorticity} shows a visualization of such structures for the same turbine presented in Section \ref{sec:simulations}, but using a finer discretization and a smaller domain. Future stages of our project will perfect these preliminary simulations, by introducing more complex models than the Lagrangian actuator disc models presented in this work.

%======================================================

\section*{Acknowledgments}
The authors thank Philippe Drobinski for very fruitful discussions related to modeling aspects of the atmospheric boundary layer.

%=====================================================
\appendix
\section{Appendix}

\subsection*{On the partial exponential Euler scheme for partially linear SDEs}\label{subsec:exposcheme}

\newcommand{\oZ}{\overline{Z}}
\newcommand{\ooZ}{\overline{\overline{Z}}}
\newcommand{\e}{{\rm e}}
\newcommand{\Dt}{\Delta t}
\newcommand{\E}{\mathcal{E}}

A part of the drift of the  fluid-particle velocity equation \eqref{eq:generic-U} is a linear term of the form 
$$-G(t,\Xx_t)(\Uu_t -\av{\Uu}(t,\Xx_t)).$$
It means that during a time step $\Delta t$, each component of the velocity behaves like a one-dimensional Ornstein-Uhlenbeck  process of generic  form:  given $Z_t$, 
\begin{align}\label{eq:OU}
dZ_s = \left(\alpha (Z_s -m) +\beta \right) ds +\sigma dW_s,~\mbox{for }s\in [t, t+\Delta t], 
\end{align}
where here we assume that  \(\alpha\le 0\), $\sigma$, $m$ and \(\beta\) are constants, or  coefficients values frozen at the time $t$. Typically, from the tensor terms $G(t,\Xx_t)$, we put in $\alpha$ only the diagonal terms, while extra diagonal terms are included in $\beta$. 

To simplify the presentation, assume that we work coordinate by coordinate, and that Brownian motion $W$ and all the coefficients are valued in $\er$. Such linear SDE \eqref{eq:OU} has an explicit solution $Z$ valued in $\er$ and   given by 
\begin{align*}
Z_{t + \Delta t}  & = \e^{\alpha \Delta t}Z_{t} 
  + \left( m -\frac{\beta}{\alpha}\right) \times \left(1-\e^{\alpha\;\Delta t}\right)
  +\sigma  \e^{\alpha\;\Delta t}\int_{t}^{t + \Delta t} \e^{-\alpha r} dW_r.
\end{align*}
It is classical to notice that $\sigma  \e^{\alpha\;t}\int_t^{t+ \Delta t} \e^{-\alpha r} dW_r$ is normally distributed with  Gaussian law $\NN(0,\gamma^2)$, where 
\[\gamma^2 = \sigma^2 \e^{2\alpha \Delta t}\int_0^{\Delta t} \e^{-2\alpha r} dr =    \frac{\sigma}{2\alpha}\left[\e^{2 \alpha \Delta t} -1\right] .\] 
We thus have the following exact   simulation formula for $Z_{t_{n+1}}$,  given $Z_{t_n}$
\begin{align*}
Z_{t_{n+1}}  = \e^{\alpha \Delta t }\left( Z_{t_n} -m + \frac{\beta}{\alpha}\right) 
  + \left( m -\frac{\beta}{\alpha}\right) 
  + \sigma \frac{\sqrt{1-\exp(2\alpha \Delta t)}}{\sqrt{ (- 2 \alpha) }} \eta_n, 
\end{align*}
where $(\eta_n)$ is a  sequence of independent  $\NN(0,1)$-identically distributed random variables. This procedure delivers a discrete time random process with the exact law of the solution of \eqref{eq:OU}, as long as $m$, $\alpha$, $\beta$ and $\sigma$ are constant.  
In the situation of Equation \eqref{eq:generic-U}, all those parameters  may correspond to frozen coefficients $\alpha_n,\sigma_n,m_n,\beta_n$ during the integration step $[t_n , t_n + \Delta t]$, and the exponential Euler scheme below becomes an approximation procedure:
\begin{align}\label{generic-expo-scheme}
\oZ_{t_{n+1}}  = \e^{\alpha_n \Delta t }\left( \oZ_{t_n} -m_n + \frac{\beta_n}{\alpha_n}\right) 
  + \left( m_n -\frac{\beta_n}{\alpha_n}\right) 
  + \sigma_n \frac{\sqrt{1-\exp(2\alpha_n \Delta t)}}{\sqrt{ (- 2 \alpha_n) }} \eta_n.
\end{align}

Such exponential scheme strategy was previously considered in the context of Lagrangian two-phase flow modeling by Minier, Peirano  and  Chibbaro \cite{minier-etal-03},  who  numerically analyses the ability  of such method to be insensitive to some limit value of the time scale $\tfrac{1}{\alpha_n}$.  

Here, our concern slightly differ. To be precise, the main advantage of the partial exponential Euler scheme to the classical Euler scheme, defined as (with the same notation for coefficients)
\begin{align*}
\ooZ_{t_{n+1}}  = \ooZ_{t_n} + \alpha_n (\ooZ_{t_n} - m_n) \Delta t + \sigma_n \sqrt{\Delta t} \eta_n, 
\end{align*}
arises from the fact that the rate of strong convergence of the exponential scheme depends only on the time variation of the map $t\mapsto \alpha(t)$ withing its Lipschitz or H\"{o}lder regularity parameters, but does not depend directly on the  values of $\alpha_n$. This is not true  for the   classical Euler scheme (see the computation below). 

Concerning the time discretisation part, the convergence of our numerical scheme, described in section \ref{sec:scheme},  is  mainly driven by the notion of the weak convergence: 
$$|\Ee f(Z_T) - \Ee f(\oZ_{T})|  \longrightarrow 0, \mbox{when  } \Delta t  \rightarrow 0,$$ 
given a set on test function $f$. But the notion of strong convergence of a $p$-th moment approximation: 
$$\left(\Ee [|Z_T - \oZ_{T}|^p]  \right)^{1/p}  \longrightarrow 0, \mbox{when  } \Delta t  \rightarrow 0$$ plays also an important role when one mix a SDE time discretisation scheme  with a Monte Carlo procedure. Indeed, a particular attention should be paid to  the upper-bound of the overall  variance of the  error: when 
$$\Ee f(Z_T) ~ \mbox{is approximated by }~ \frac{1}{N} \sum _{i=1}^N f(\oZ_{T}^{(i)}), $$ 
 classical error analysis decomposes the error in the time discretisation error part (bias bound)  and the mean square Monte Carlo error part  
$$|\Ee f(Z_T) - \Ee f(\oZ_{T})|   
+ \sqrt{\Ee\left[\Big(\Ee f(\oZ_{T}) -  \frac{1}{N} \sum _{i=1}^N f(\oZ_{T}^{(i)})\Big)^2\right]}.$$
It is well known that the variance of the error
$$\Var\Big( \Ee f(Z_{T}) -  \frac{1}{N} \sum _{i=1}^N f(\oZ_{T}^{(i)}) \Big)$$
can be  control with the strong error (as soon as $f$ is  for instance Lipschitz, see e.g Kebaier \cite{kebaier2005statistical}).  Thus, in our case the exponential scheme prevents the  variance of the error to fluctuate too much with huge values of  \(\vert\alpha_n\vert\).  
Let us now recall that in our Lagrangian stochastic model, $\alpha_n$ is mainly the square root of the 
turbulent kinetic energy $\tke$  in  \eqref{eq:lagrang_tke_sdm}, computed itself with the PIC estimator \eqref{eq:numesp} for the variance of the velocity components. It is then numerically pertinent to prefer the partial exponential scheme in order to stabilize the variance of $\tke$  estimator. 

\medskip

Notice that all this empirical analysis is contingent to the fact that $(\alpha_n,n\in\nat)$ stays negative. 

\medskip
For the sake of completeness, we detail below some basic computations in order to compare the behavior of the Euler schemes and Exponential Euler schemes on the generic linear SDE 
\begin{align}\label{eq:OUtime}
dZ_t = \left(\alpha(t)  (Z_t - m(t) ) +\beta(t) \right) dt +\sigma(t) dW_t,\quad Z_0 \mbox{ given. }
\end{align}
\subsubsection*{On the moments stability and the strong convergence of the discussed  schemes for the SDE   \eqref{eq:OUtime}}  
Assume that the coefficients $\alpha(\cdot)$, $m(\cdot)$,  $\beta(\cdot)$, $\sigma(\cdot)$ in \eqref{eq:OUtime} are continuous on $[0,T]$ for a finite final time $T$. 
As an affine SDE \eqref{eq:OUtime} has a unique strong solution.  Using Itô formula, it is easy to check that the solution is given by the following closed expression 
\begin{align}\label{eq:OUtime=sol}
Z_t = &Z_0 \e^{\int_0^t \alpha(r) dr} -  \int_0^t \e^{\int_s^t \alpha(r) dr} \alpha(s) m(s) ds  + \int_0^t \e^{ \int_s^t \alpha(r) dr} \beta(s) ds  + \int_0^t \e^{\int_s^t \alpha(r) dr}\sigma(s) dW_s  
\end{align}
Freezing the coefficients on each subinterval $[n\Delta t, (n+1)\Delta t)$,  we consider the continuous version of the exponential scheme 
\begin{align}\label{eq:OU-ESC}
\begin{aligned}
\oZ_t = &Z_0 \e^{\int_0^t \alpha(\tau(s) ) dr}  -  \int_0^t  \e^{\int_s^t \alpha(\tau(r))  dr} \alpha(\tau(s)) m((\tau(s)) ds\\
& + \int_0^t \e^{\int_s^t \alpha((\tau(r)) dr} \beta((\tau(s)) ds + \int_0^t \e^{\int_s^t \alpha((\tau(r)) dr} \sigma((\tau(s)) dW_s  
\end{aligned}
\end{align}
where $\tau(t)=\sup_{k\in\{1,\ldots,N\}}\{t_k:t_k\leq t\}$, which coincide with the definition 
\begin{align*}
\oZ_{t_{n +1}}  & = \oZ_{t_n}  \e^{\Dt \alpha(t_n)}   -  \int_{t_n}^{t_{n +1}}  \e^{(t_{n +1} -s) \alpha(t_n)}  \alpha(t_n)  m(t_n)  ds \\
& \quad +     \int_{t_n}^{t_{n +1}}  \e^{(t_{n +1} -s) \alpha(t_n)} \beta(t_n)  ds +    \int_{t_n}^{t_{n +1}} \e^{(t_{n +1} -s) \alpha(t_n)} \sigma(t_n) dW_s
\end{align*}
at each time step $t_n$. 
Using the closed forms, it is straightforward to compute a bound of any even $2p$-th moment $p\geq1$, using successively the  Jensen Inequality and Itô Isometry
\begin{align*}
\Ee[Z_t^{2p}] \leq  & 4^{2p-1} \Ee[ (Z_0^{2p} \e^{2p \int_0^t \alpha(r) dr}] + \Ee\left[\left( \int_0^t \e^{\int_s^t \alpha(r) dr}\alpha(s) m(s) ds\right)^{2p} \right]  \\
& + \Ee\left[\left(\int_0^t  \e^{\int_s^t \alpha(r) dr} \beta(s) ds\right)^{2p}  \right]   +\Ee \left[\left( \int_0^t \e^{2 \int_s^t \alpha(r) dr}\sigma(s)^2  ds  \right)^{p} \right].
\end{align*}
Assuming in addition that $\alpha$ is valued in $(-\infty, 0]$, we easily get that $\Ee[Z_t^{2p}] $ is bounded by a constant uniform in $\alpha$. This is also true for $\Ee[\oZ_t^{2p}] $ for the same reason.  From a different computation, this stability property is also true for the classical Euler scheme 
\begin{align*}
\ooZ_{t_{n +1}}   = \ooZ_{t_n}  +  \int_{t_n}^{t_{n +1}}  \alpha(t_n) ( \ooZ_{t_n} - m(t_n) )  ds +     \int_{t_n}^{t_{n +1}}  \beta(t_n)  ds +    \int_{t_n}^{t_{n +1}} \sigma(t_n) dW_s
\end{align*}
whose continuous version is 
\begin{align}\label{eq:OU-EuSC}
\begin{aligned}
\ooZ_t = &Z_0 +  \int_0^t \alpha(\tau(s)) ( \ooZ_{\tau(s)} - m(\tau(s)) )  ds +     \int_0^t \beta(\tau(s))  ds +    \int_0^t \sigma(\tau(s) dW_s
\end{aligned}
\end{align}
Indeed,  applying the Itô formula to $\ooZ_t^{2p}$ and classical computations 
\begin{align*}
\Ee[\ooZ_t^{2p}]   & =   \Ee[Z_0^{2p}] \e^{\int_0^t 2p \alpha(s) ds} 
 + \int_0^t 2p \e^{\int_s^t 2 p \alpha(s) ds)} \left\{ \beta(s)  -   \alpha(s)  m(s)\right\} \Ee[\ooZ^{2p-1}_s]   ds \\ 
& \quad + \int_0^t 2p (2p-1)   \e^{\int_s^t 2 p \alpha(s) ds}  \frac{\sigma(s)^2}{2} \Ee[\ooZ_s ^{2p-2}] ds  
\end{align*}
and again, due to the sign of $\alpha$,  we easily conclude. 

To simplify the presentation, we consider now the trajectorial error $t\mapsto \left(\Ee[|Z_t - \oZ_t|^{2p}]\right)^{1/2p}$ when other coefficients $m$, $\beta$ and $\sigma$ are some constants.  So $Z_t  - \oZ_t $ writes 
\begin{align*}
Z_t  - \oZ_t 
 & = \left(Z_0 - m\right)  \left(\e^{\int_0^t \alpha(s) ds} - \e^{\int_0^t \alpha(\tau(s) ) ds}\right)  + \beta \int_0^t \left( \e^{\int_s^t \alpha((\tau(r)) dr}  - \e^{\int_s^t \alpha((\tau(r)) dr} \right) ds \\
&\quad  +\sigma \int_0^t\left(  \e^{\int_s^t \alpha(r) dr}  -  \e^{\int_s^t \alpha((\tau(r)) dr}\right)  dW_s.
\end{align*}
The strong error bound  is then derived from the bound of terms of the form 
$\E_t := \left(\e^{\int_0^t \alpha(s) ds} - \e^{\int_0^t \alpha(\tau(s) ) ds}\right)$, that also writes 
\begin{align*}
\E_t  = \e^{\int_0^t \alpha(\tau(s))ds}  \int_0^t \Big(\alpha(s) - \alpha(\tau(s)\Big) \e^{\int_0^s \alpha(r)dr} \e^{ -  \int_0^s \alpha(\tau(r))dr }   ds. 
\end{align*}
This last expression leads to  the upper bound $|\E_t| \leq \int_0^t |\alpha(s) - \alpha(\tau(s))|  \leq L \Dt $, when for instance $\alpha$  is Lispschitz with constant $L$. Using again  Jensen Inequality and Itô Isometry,   we get that $\sup_{t\in[0,T]} \Ee[|Z_t - \oZ_t|^{2p}]\leq C |\E_t |^{p}$  where $C$ does not depend on $\alpha$. 
In particular, when $\alpha$ is also a constant, the strong error vanishes and we recover the fact that the exponential scheme is exact in that case. 

\medskip
The strong convergence for  the Euler scheme differs on this last point,  even when all the coefficients are some constant. Indeed, starting with the It{\^{o}} formula, we have the following 
\begin{align*}
\Ee[(Z_t -\ooZ_t)^{2p}] & =  \int_0^t 2p \alpha \Ee[(Z_s - \ooZ_s)^{2p-1} (Z_s  - \ooZ_{\tau(s)})] ds \\
& = \int_0^t 2p \alpha \Ee[(Z_s - \ooZ_s)^{2p}]  ds  + \int_0^t 2p  \alpha\Ee[ (Z_s - \ooZ_s)^{2p-1}  (\ooZ_s  - \ooZ_{\tau(s)}) ] ds. 
\end{align*} 
From that point, if  $\alpha$ is negative,  the first term in the right hand side disappears, but the second one, multiplied by $\alpha$ is not a signed term, so any Young or H{\"o}lder inequality to separate this second term in what is called the  local error $\Ee[ (\ooZ_s  - \ooZ_{\tau(s)})^{2p} ]$ with the iteration term $ \Ee[ (Z_s - \ooZ_s)^{2p}]] $ will make appear a $|\alpha|$ in the final bound. For instance, with  Young inequality and next  Gr{\"o}nwall lemma
\begin{align*}
\Ee[(Z_t -\ooZ_t)^{2p}] \leq    \int_0^t (2p-1) |\alpha|  \Ee[ (Z_s - \ooZ_s)^{2p}]] ds + \int_0^t  |\alpha|  \Ee[ (\ooZ_s  - \ooZ_{\tau(s)})^{2p} ] ds \leq C \Delta t^p
\end{align*} 
with a constant $C$ that depends on $|\alpha|$. 

\bibliographystyle{plain}

\bibliography{biblio}

%=====================================================

\end{document}